\documentclass[fleqn,usenatbib,twocolumn]{mnras}
\usepackage{newtxtext,newtxmath}
\usepackage[T1]{fontenc}
\usepackage{ae,aecompl}
\usepackage{xspace}
\usepackage{graphicx}
\usepackage{amsmath}
\usepackage{multicol}
\usepackage{wrapfig}
\usepackage{threeparttable}
\usepackage[normalem]{ulem}

\newcommand\chisq{\chi^{2}}

\newcommand\Mea{\mathrm{M_{\oplus}}}
\newcommand\Rea{\mathrm{R_{\oplus}}}

\newcommand\Rjup{\mathrm{R_\mathrm{Jup}}}

\newcommand\Msun{\mathrm{M_{\odot}}}
\newcommand\Rsun{\mathrm{R_{\odot}}}

\newcommand\microm{\mathrm{\mu m}}
\newcommand\mps{\mathrm{m\,s^{-1}}}

\newcommand\kmps{\mathrm{km\,s^{-1}}}
\newcommand\mpspd{\mathrm{m\,s^{-1}\,day^{-1}}}

\newcommand\gpcmcmcm{\mathrm{g\,cm^{-3}}}

\newcommand\maspyr{\mathrm{mas\,yr^{-1}}}
\newcommand{\splx}{$\pi_{\star}$} 
\newcommand{\sdist}{$d_{\star}$}
\newcommand\steff{T_\mathrm{eff}}
\newcommand\logsg{\log\,g_{\star}}
\newcommand\sfeh{\mathrm{[Fe/H]}}

\newcommand\sm{M_{\star}}
\newcommand\sr{R_{\star}}
\newcommand\sden{\rho_{\star}}

\newcommand\spmra{\mathrm{PM_\mathrm{RA}}}
\newcommand\spmdec{\mathrm{PM_\mathrm{DEC}}}

\newcommand\svmic{\mathit{v}_\mathrm{mic}}
\newcommand\svmac{\mathit{v}_\mathrm{mac}}

\newcommand\svrotsini{\mathit{v}_\mathrm{rot} \sin\,i_{\star}}

\newcommand\sprot{P_\mathrm{rot}}

\newcommand\ppr{R_\mathrm{p}}
\newcommand\ppden{\rho_\mathrm{p}}

\newcommand\pbm{M_\mathrm{b}}

\newcommand\pbr{R_\mathrm{b}}
\newcommand\pbden{\rho_\mathrm{b}}

\newcommand\gaia{\emph{\it Gaia}}

\newcommand\kepler{\emph{{\it Kepler}}}
\newcommand\ktwo{\emph{{\it K2}}}
\newcommand\tess{\emph{{\it TESS}}}

\newcommand{\host}{K2-280}
\newcommand{\planet}{K2-280\,b}
\newcommand{\splxv}[1][mas]{${2.526}\,{\pm}\,{0.111}$~#1}
\newcommand{\sdistv}[1][pc]{$391.5^{+7.5}_{-7.2}$~#1}

\newcommand{\sfehv}[1][$\mathrm{dex}$]{${0.33}\,{\pm}\,{0.08}$~#1}

\newcommand{\smass}[1][$M_{\odot}$]{ $ 1.03 \pm 0.03 $ #1} 
\newcommand{\sradius}[1][$R_{\odot}$]{ $1.28 \pm 0.07 $ #1}
\newcommand{\stemp}[1][$\mathrm{K}$]{ $ 5500 \pm 100 $ #1 }
\newcommand{\Tzerob}[1][days]   {$7307.58114 \pm 0.00056$~#1} 
\newcommand{\Pb}[1][days]   {$19.89526  \pm 0.00028$~#1} 
\newcommand{\esinb}[1][ ]   {$-0.547 _{ - 0.05 } ^ { + 0.047 }$~#1} 
\newcommand{\ecosb}[1][ ]   {$-0.235 _{ - 0.043 } ^ { + 0.038 }$~#1} 
\newcommand{\bb}[1][ ]   {$0.27 _{ - 0.17 } ^ { + 0.16 }$~#1} 
\newcommand{\arb}[1][ ]   {$25.79 _{ - 0.90} ^ { + 0.87 }$~#1} 
\newcommand{\rrb}[1][ ]   {$0.05354 _{ - 0.00056 } ^ { + 0.00094 }$~#1} 
\newcommand{\kb}[1][${\rm m\,s^{-1}}$]   {$9.18 \pm 1.27$~#1} 
\newcommand{\mpb}[1][$M_{\oplus}$]{$37.1 \pm 5.6$~#1} 
\newcommand{\rpb}[1][$R_{\oplus}$]{$7.50 \pm 0.44$~#1} 
\newcommand{\Tperib}[1][days]   {$7315.06 _{ - 0.44 } ^ { + 0.38 }$~#1} 
\newcommand{\eb}[1][ ]{$0.35_{-0.04}^{+0.05}$~#1} 
\newcommand{\wb}[1][deg]   {$246.77_{-5.28 }^{+4.54}$~#1}
 
\newcommand{\ab}[1][AU]   {$0.1461 _{ - 0.0097 } ^ { + 0.0099 }$~#1}

\newcommand{\densspb}[1][${\rm g\,cm^{-3}}$]   {$0.8 _{ - 0.13 } ^ { + 0.16 }$~#1} 
\newcommand{\denpb}[1][${\rm g\,cm^{-3}}$]   {$0.48 _{ - 0.10 } ^ { + 0.13 }$~#1} 
\newcommand{\grapb}[1][${\rm cm\,s^{-2}}$]   {$648.0 _{ - 102.0 } ^ { + 107.0 }$~#1} 
\newcommand{\grapparsb}[1][${\rm cm\,s^{-2}}$]   {$647.0 _{ -117.0 } ^ { + 133.0 }$~#1} 
\newcommand{\Teqb}[1][K]   {$787 \pm 17$~#1} 
\newcommand{\ttotb}[1][hours]   {$8.267 _{ - 0.054 } ^ { + 0.063 }$~#1}

\newcommand{\qone}[1][]   {$0.54 _{ - 0.11 } ^ { + 0.14 }$~#1} 
\newcommand{\qtwo}[1][]   {$0.249 _{ - 0.071 } ^ { + 0.082 }$~#1} 
\newcommand{\uone}[1][]   {$0.367 _{ - 0.079 } ^ { + 0.074 }$~#1} 
\newcommand{\utwo}[1][]   {$0.37 _{ - 0.15 } ^ { + 0.15 }$~#1} 
\newcommand{\HARPSN}[1][${\rm km\,s^{-1}}$]   {$-1.1934 _{ - 0.0011 } ^ { + 0.0011 }$~#1} 
\newcommand{\HARPS}[1][${\rm km\,s^{-1}}$]   {$-1.1907 _{ - 0.0011 } ^ { + 0.0011 }$~#1} 
\newcommand{\FIES}[1][${\rm km\,s^{-1}}$]   {$-1.2349 _{ - 0.0039 } ^ { + 0.0038 }$~#1} 
\newcommand{\jHARPSN}[1][${\rm m\,s^{-1}}$]   {$0.36 _{ - 0.3 } ^ { + 0.97 }$~#1} 
\newcommand{\jHARPS}[1][${\rm m\,s^{-1}}$]   {$1.28 _{ - 1.13 } ^ { + 1.76 }$~#1} 
\newcommand{\jFIES}[1][${\rm m\,s^{-1}}$]   {$0.67 _{ - 0.58 } ^ { + 2.58 }$~#1}

\title[\planet\ --- a low density warm sub-Saturn around a mildly evolved star]{\planet\ - a low density warm sub-Saturn around a mildly evolved star}

\author[G. Nowak et al.]{Grzegorz Nowak$^{1,2}$\thanks{E-mail: gnowak@iac.es},
Enric~Palle$^{1,2}$,
Davide~Gandolfi$^{3}$,
Hans~J.~Deeg$^{1,2}$, 
\newauthor
Teruyuki~Hirano$^{4}$,
Oscar~Barrag\'{a}n$^{5,3}$, 
Masayuki~Kuzuhara$^{6,7}$,
Fei~Dai$^{8,9}$,
\newauthor
Rafael~Luque$^{1,2}$,
Carina~M.~Persson$^{10}$,
Malcolm~Fridlund$^{10,11}$, 
\newauthor
Marshall~C.~Johnson$^{12}$,
Judith~Korth$^{13}$,
John~H.~Livingston$^{14}$,
Sascha~Grziwa$^{13}$,
\newauthor
Savita~Mathur$^{1,2}$,
Artie~P.~Hatzes$^{15}$,
Jorge~Prieto-Arranz$^{1,2}$,
David~Nespral$^{1,2}$, 
\newauthor
Diego~Hidalgo$^{1,2}$,
Maria~Hjorth$^{16}$,
Simon~Albrecht$^{16}$, 
Vincent~Van~Eylen$^{8}$,
\newauthor
Kristine~W.~F.~Lam$^{17}$,
William~D.~Cochran$^{18}$,
Massimiliano~Esposito$^{15}$,
\newauthor
Szil\'ard~Csizmadia$^{19}$, 
Eike~W.~Guenther$^{15}$,
Petr~Kabath$^{20}$,
Pere~Blay$^{21}$,
\newauthor
Rafael Brahm$^{22,23,24}$,
Andr\'{e}s Jord\'{a}n$^{25,23,24}$,
N\'{e}stor Espinoza$^{26}$,
Felipe Rojas$^{24,23}$,
\newauthor
N\'{u}ria~Casasayas~Barris$^{1,2}$,
Florian~Rodler$^{27}$,
Roi~Alonso~Sobrino$^{1,2}$,
Juan~Cabrera$^{19}$,
\newauthor
Ilaria~Carleo$^{28}$,
Alexander~Chaushev$^{17}$,
Jerome~de~Leon$^{14}$,
Philipp~Eigm\"uller$^{19}$,
\newauthor
Michael~Endl$^{18}$,
Anders~Erikson$^{19}$,
Akihiko~Fukui$^{7}$,
Iskra~Georgieva$^{10}$,
\newauthor
Luc\'ia~Gonz\'alez-Cuesta$^{1,2}$,
Emil~Knudstrup$^{16}$,
Mikkel~N.~Lund$^{16}$,
\newauthor
Pilar~Monta\~nes Rodr\'iguez$^{1,2}$,
Felipe~Murgas$^{1,2}$,
Norio~Narita$^{6,29,7,1}$,
\newauthor
Prajwal~Niraula$^{30}$,
Martin~P\"atzold$^{13}$,
Heike~Rauer$^{19,17}$,
Seth~Redfield$^{28}$,
\newauthor
Ignasi~Ribas$^{31,32}$,
Marek~Skarka$^{20,33}$,
Alexis~M.~S.~Smith$^{19}$,
and Jano~Subjak$^{20,34}$\\
\\
Affiliations are listed at the end of the paper
}
\date{Accepted 2020 July 8. Received 2020 July 7; in original form 2019 July 8}
\pubyear{2019}

\begin{document}
\label{firstpage}
\pagerange{\pageref{firstpage}--\pageref{lastpage}}
\maketitle

\begin{abstract}
We present an independent discovery and detailed characterisation of \planet, a transiting low density warm sub-Saturn in a 19.9-day moderately eccentric orbit ($e$\,=\,\eb) from \ktwo\ campaign 7. A joint analysis of high precision HARPS, HARPS-N, and FIES radial velocity measurements and \ktwo\ photometric data indicates that \planet\ has a radius of $\pbr$\,=\,\rpb\ and a mass of $\pbm$\,=\,\mpb, yielding a mean density of $\pbden$\,=\,\denpb. The host star is a mildly evolved G7 star with an effective temperature of $\steff$\,=\,\stemp, a surface gravity of $\logsg$\,=\,4.21\,$\pm$\,0.05 (cgs), and an iron abundance of $\sfeh$\,=\,\sfehv, and with an inferred mass of $\sm$\,=\,\smass\ and a radius of $\sr$\,=\,\sradius. We discuss the importance of \planet\ for testing formation scenarios of sub-Saturn planets and the current sample of this intriguing group of planets that are absent in the Solar System.
\end{abstract}


\section{Introduction}
The main advantage of the extended NASA's \kepler\ mission \citep{2010Sci...327..977B}, known as \ktwo\ \citep{2014PASP..126..398H}, was a much larger number of bright stars in its fields of view located along the ecliptic. A significant number of planets transiting bright stars have been discovered in all \ktwo\ campaigns \citep[e.g.,][]{2015ApJ...809...25M,2016ApJS..226....7C,2016ApJS..222...14V,2017AJ....154..207D,2018AJ....155...21P,2018AJ....155..136M,2018AJ....156...22Y,2018ApJS..239....5C,2018AJ....156..277L}. Some of these planets were only validated, but many were characterized by means of high precision radial velocity (RV) measurements that enabled mass determination with precision better than 20\% \citep[e.g.,][]{2015ApJ...800...59V, 2017AJ....154..123G,2017AJ....154..122C,2018A&A...618A.116P,2018AJ....155...72R,2018A&A...612A..95B,2018AJ....155..107M}. However much higher precision is needed to distinguish between various possible planetary compositions \citep[see e.g.][and references therein]{2015A&A...577A..83D}. Mass determination with such precision for small planets ($R_\mathrm{p}$\,=\,1--4\,$\Rea$) were possible only for short period ($P_\mathrm{orb}$\,$\lesssim$\,10\,days) sub-Neptunes and super-Earths, which induces RV semi-amplitudes on the parent stars of a few $\mps$, and for stars hosting ultra-short period planets, for which the Doppler reflex motion is enhanced by the extremely short orbital period  \citep[$P_\mathrm{orb}$\,$<$\,1\,day; ][and references therein]{2018NewAR..83...37W}. Most of the small \ktwo\ planets with precise mass determination orbit bright stars, i.e., stars brighter than $V$\,=\,11.5, which is the current limit for $\sim$1\,$\mps$ precision with spectrographs mounted at 3--4-m class telescopes \citep{2013Natur.503..377P}. The constraints for precise determination of planetary masses are naturally more relaxed for higher-mass planets, enabling us to study super-Neptune/sub-Saturn planets ($R_\mathrm{p}$\,=\,4--8\,$\Rea$) with longer orbital periods around fainter stars.

Sub-Saturns form a very intriguing group of planets that have no counterpart in the Solar System. Their main characteristic is a significant contribution of both heavy metal cores and low density gaseous envelopes to the total planet mass \citep{2016ApJ...818...36P}. They are thus important laboratories to study envelope accretion. As shown by \cite{2017AJ....153..142P}, the population of sub-Saturns has a very uniform distribution of the planetary mass between $\sim$6 and 60~$\Mea$. Similar to gas-giant planets, they are also found to orbit mainly metal-rich stars. Finally, the most massive sub-Saturns are often the only detected planet in the system and orbit their parent stars on eccentric orbits, which suggests that dynamical instability might have played an important role in their formation.

Here we present an independent discovery and characterisation of a low mass, sub-Saturn planet on a 20-d orbit around a relatively faint ($V$\,=\,12.5), metal rich ($\sfeh$\,=\,\sfehv), slightly evolved \ktwo\ star that was proposed as a planet candidate by \cite{2018AJ....155...21P} and \cite{2018AJ....155..136M}, and statistically validated as a planet by \cite{2018AJ....156..277L}. This kind of planets was usually avoided by RV follow-up of \ktwo\ candidates because of the faintness of their host stars. Besides, slightly evolved stars are typically avoided in RV follow-up projects, because of their higher expected stellar jitter \citep[see e.g.][and references therein]{2006A&A...454..943H,2008A&A...480..215H,2019ApJ...883..195T}. Both these effects may bias the statistical analysis of warm-giant planets. \planet\ joins a sample of 30 sub-Saturns with mean densities determined with precision better than 50\% discovered mainly by \kepler\ and \ktwo\ \citep[see][and references therein for first 23 planets]{2017AJ....153..142P}.

This work was done as a part of the KESPRINT collaboration\footnote{\url{https://www.iac.es/proyecto/kesprint/}.}, which aims to confirm and characterise \ktwo\ and \tess\ planets. In Section\,\ref{sec-observations} we describe the observations of \host, specifically the \ktwo\ photometry, the NOT/FIES, ESO/HARPS, and TNG/HARPS-N high-resolution spectroscopy follow-up, and the high-contrast imaging. In Section~\ref{sec-host_star} and \ref{sec-global_analysis} we present the properties of the host star \host\ and the global analysis of photometric and Doppler data, respectively. In Section~\ref{sec-disasum} we finally summarise and discuss the characteristics of \planet\ in the context of the properties of the known population of sub-Saturn planets with mean densities determined with precision better than 50\%.

\section{Observations and data reductions}
\label{sec-observations}

\subsection{\ktwo\ Photometry}
\label{subsec-observations-k2}

\begin{figure}
\centerline{\includegraphics[angle=0,width=\linewidth]{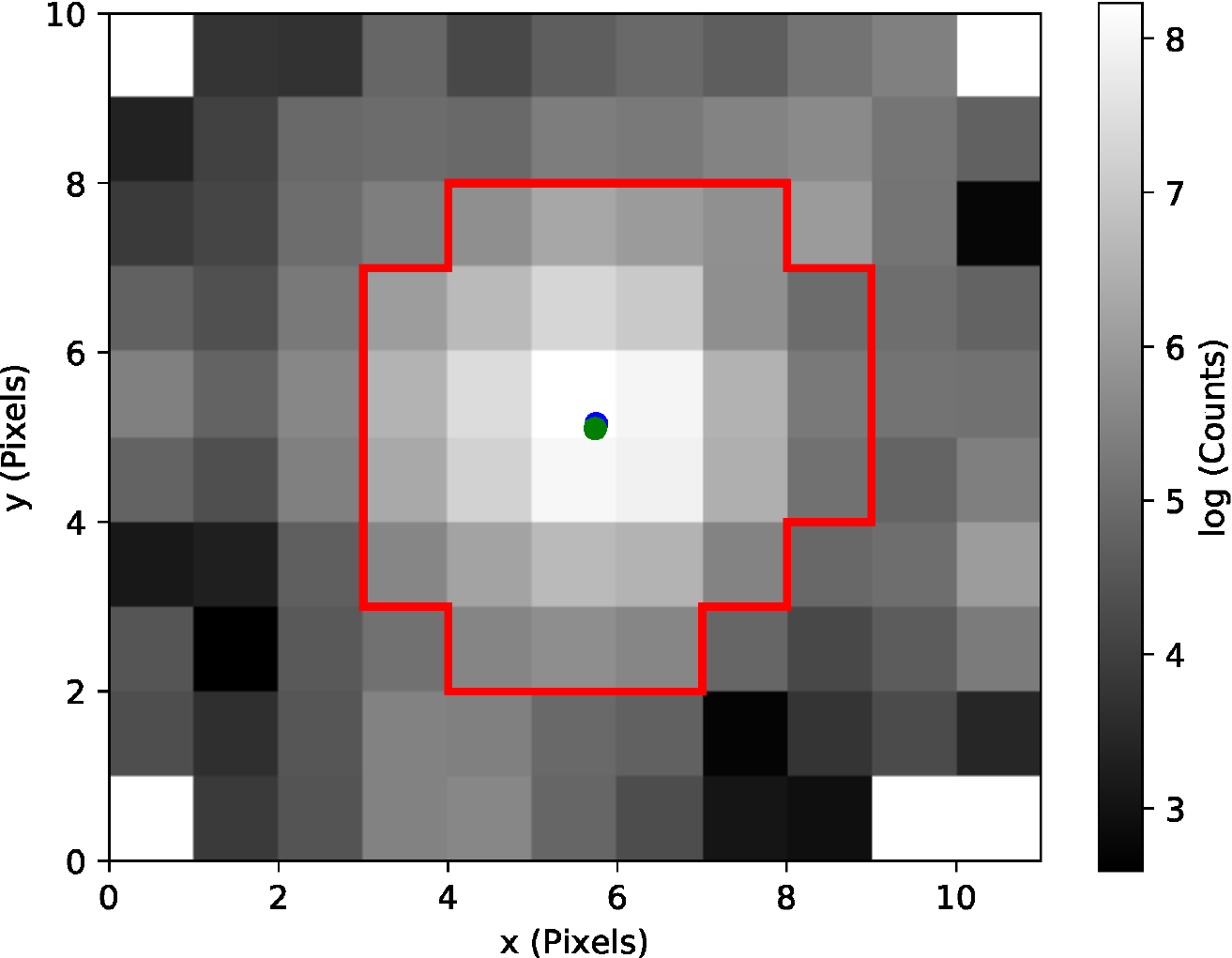}}
\caption{\ktwo\ image of \host. Red lines shows the aperture defined by the amount of light of each pixel and level of background light. The electron count is indicates by the intensity of shading (light grey for high and dark grey for low count). The green circle indicates the current position of the target in the EPIC catalog, and the blue circle is the centre of the flux distribution. The scale of the image is the \kepler\ pixel scale of 3.98 arcsec/pix. The K2 image is not a rectangle, but it is irregularly shaped. The white pixels in the corners contain no data.
\label{figure-01}}
\end{figure}

\begin{figure}
\centerline{\includegraphics[angle=0,scale=0.125]{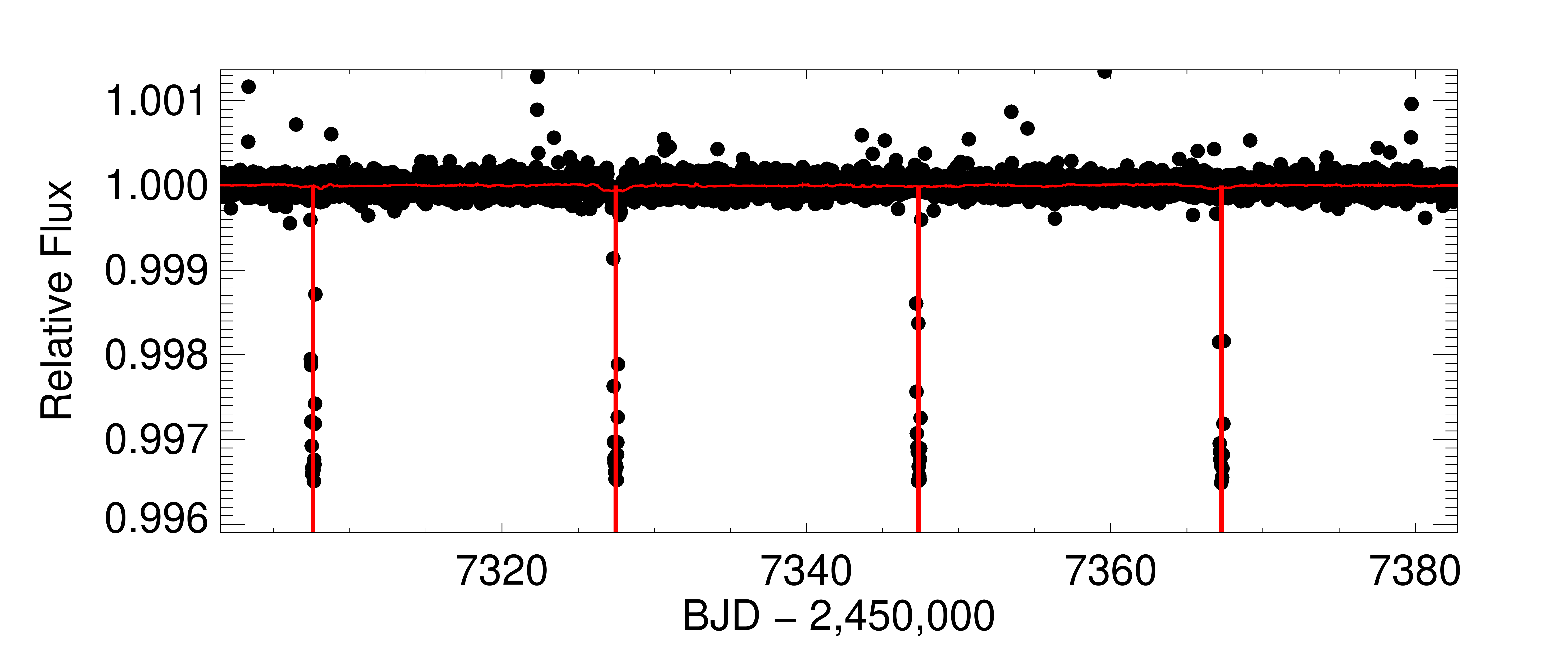}}
\caption{Detrended \ktwo\ light curve of \host. The 4 transits observed by \ktwo\ are marked by vertical solid red lines. The horizontal red line is the local median flux level with a window of 0.5-day.
\label{figure-02}}
\end{figure}

\host\ was one of 13\,469 long cadence targets observed from October 4$^{\mathrm{th}}$ to December 26$^{\mathrm{th}}$ 2015 (UT) during \ktwo\ campaign 7. It was proposed as a target by GO programmes 7030 (PI Howard), and 7085 (PI Burke). We downloaded \host\ images from the MAST archive\footnote{\url{https://archive.stsci.edu/k2/data\_search/search.php}.} and used them to produce a de-trended \ktwo\ light curve as described in detail in \cite{2017AJ....153...40D}. Figure~\ref{figure-01} shows the pixel mask used to perform simple aperture photometry. We used the box fitting least-square (BLS) routine \citep{2002A&A...391..369K,2010ApJ...713L..87J}, improved by implementation of the optimal frequency sampling described in \cite{2014A&A...561A.138O} to search for transiting planet candidates in all Field 7 targets light curves. We detected transits of \planet\ with a signal-to-noise ratio (SNR) of 24.5, depth of $\sim$3.5 $\times 10^{-3}$, period of $P\,=\,19.89518\,\pm\,0.00028$ days, and a mid-time of the first transit $T_{0}\,=\,2457307.58101\,\pm\,0.00059$ days in Barycentric Julian Date in the Barycentric Dynamical Time \citep[BJD$_\mathrm{TDB}$; see, e.g.,][]{2010PASP..122..935E}. The de-trended light curve of \host\ with the correction for baseline flux variations and centroid motions is presented in Figure~\ref{figure-02} with the 4 transits observed by \ktwo\ highlighted with red lines. We removed the baseline flux variation by fitting a spline function with a width of 3 days. In Table~\ref{table-01} we report the main identifiers of \host, along with its equatorial coordinates, space motion, distance, and optical and near-infrared magnitudes.

\begin{table}
\caption{Properties of \host.
\label{table-01}}
\begin{center}
\begin{tabular}{lrr}
\hline
\hline
\noalign{\smallskip}
Parameter & Value & Source\\
\noalign{\smallskip}
\hline
\noalign{\smallskip}
\multicolumn{3}{c}{\emph{Coordinates and Main Identifiers}}\\
\noalign{\smallskip}
RA 2000.0 (hours)       &  19:26:22.881         & \gaia\ DR2\\
Dec 2000.0 (deg)        & -22:14:51.552         & \gaia\ DR2\\
\gaia\ DR2 Identifier   & 6772454416893148928   & \gaia\ DR2\\
2MASS Identifier        & 19262288-2214514      & 2MASS PSC\\
UCAC Identifier         & 339-184113            & UCAC4\\
EPIC Identifier         & 216494238             & EPIC\\
TIC Identifier          & 119605900             & TIC\\
\noalign{\smallskip}
\hline
\noalign{\smallskip}
\multicolumn{3}{c}{\emph{Optical and Near-Infrared Magnitudes}}\\
\noalign{\smallskip}
$K_{p}$ (mag)   & 12.302                   & \ktwo\ EPIC\\
$B$ (mag)       & 13.269\,$\pm$\,0.010     & UCAC4\\
$V$ (mag)       & 12.536\,$\pm$\,0.040     & UCAC4\\
$R$ (mag)       & 12.41\,$\pm$\,0.07       & UCAC4\\
$G$ (mag)       & 12.3604\,$\pm$\,0.0002   & \gaia\ DR2\\
$g$ (mag)       & 12.850\,$\pm$\,0.020     & UCAC4\\
$r$ (mag)       & 12.320\,$\pm$\,0.020     & UCAC4\\
$i$ (mag)       & 12.067\,$\pm$\,0.040     & UCAC4\\
$J$ (mag)       & 11.141\,$\pm$\,0.021     & 2MASS\\
$H$ (mag)       & 10.854\,$\pm$\,0.024     & 2MASS\\
$K$ (mag)       & 10.765\,$\pm$\,0.019     & 2MASS\\
\noalign{\smallskip}
\hline
\noalign{\smallskip}
\multicolumn{3}{c}{\emph{Space Motion and Distance}}\\
\noalign{\smallskip}
$\spmra$ ($\maspyr$)                       & 4.44\,$\pm$\,0.08     & \gaia\ DR2\\
$\spmdec$ ($\maspyr$)                      & -12.50\,$\pm$\,0.07   & \gaia\ DR2\\
RV$_{\gamma,\mathrm{HARPS}}$ ($\kmps$)     & \HARPSN[]             & This work\\
RV$_{\gamma,\mathrm{HARPS-N}}$ ($\kmps$)   & \HARPS[]              & This work\\
RV$_{\gamma,\mathrm{FIES}}$ ($\kmps$)      & \FIES[]               & This work\\
\splx (mas)                                & \splxv[]              & \gaia\ DR2\\
\sdist (pc)                                & \sdistv[]             & \cite{2018AJ....156...58B}\\
$U$ ($\kmps$)                              & -7.78\,$\pm$\,0.06    & This work\\
$V$ ($\kmps$)                              & -5.98\,$\pm$\,0.84    & This work\\
$W$ ($\kmps$)                              & -9.18\,$\pm$\,0.71    & This work\\
\noalign{\smallskip}
\hline
\noalign{\smallskip}
\multicolumn{3}{c}{\emph{Photospheric Parameters}}\\
\noalign{\smallskip}
$\steff$ (K)     & 5500\,$\pm$\,100    & This work\\
$\logsg$$^{(a)}$ (dex)   & 4.00\,$\pm$\,0.10   & This work\\
$\logsg$$^{(b)}$ (dex)   & 4.21\,$\pm$\,0.05   & This work\\
$\sfeh$ (dex)    & 0.33\,$\pm$\,0.08   & This work\\
\noalign{\smallskip}
\hline
\noalign{\smallskip}
\multicolumn{3}{c}{\emph{Derived Physical Parameters}}\\
\noalign{\smallskip}
$\sm$ ($\Msun$)         & \smass[]                 & This work\\
$\sr$ ($\Rsun$)         & \sradius[]               & This work\\
$\sden$ ($\gpcmcmcm$)   & \densspb[]               & This work\\
Age (Gyr)               & 8.96\,$\pm$\,1.70        & This work\\
\noalign{\smallskip}
\hline
\noalign{\smallskip}
\multicolumn{3}{c}{\emph{Stellar Rotation}}\\
\noalign{\smallskip}
$\svrotsini$ ($\kmps$)   & 3.0\,$\pm$\,1.0   & This work\\
\noalign{\smallskip}
\hline
\end{tabular}
\item \emph{Notes} -- $^{(a)}$ From spectroscopy. $^{(b)}$ From stellar mass and radius. 
\end{center}
\end{table}

\subsection{High-dispersion spectroscopy}
\label{subsec-observations-hrs}
High-dispersion spectroscopic observations of \host\ were obtained between April 30$^\mathrm{th}$ 2016 (UT) and May 7$^\mathrm{th}$ 2019 (UT) using ESO/HARPS, TNG/HARPS-N, and NOT/FIES spectrographs. We collected a total of 18 HARPS, 14 HARPS-N and 6 FIES spectra. The details of these observations are given in the subsections below. Table~\ref{table-02} gives the time stamps of the spectra in BJD$_{\mathrm{TDB}}$, the RVs along with their $1\sigma$ error bars, as well as the bisector inverse slope (BIS) and full-width at half maximum (FHWM) of the cross-correlation function (CCF).

\subsubsection{ESO/HARPS}
We started the RV follow-up of \host\ using the High Accuracy Radial velocity Planet Searcher (HARPS) spectrograph \citep[R$\approx$115\,000]{2003Msngr.114...20M} mounted at the ESO 3.57-m telescope of La Silla Observatory in Chile. We acquired 18 spectra between April 30$^\mathrm{th}$ 2016 (UT) and April 27$^\mathrm{th}$ 2018 (UT) under the observing programmes 097.C-0571(B), 097.C-0948(A), 098.C-0860(A), 099.C-0491(A), 0101.C-0407(A), and 60.A-9700(G), setting the exposure times to 1200--3600 seconds. The dedicated on-line HARPS Data Reduction Software (DRS) was used to reduce the spectra, and extract the Doppler measurements and spectral activity indicators. The SNR per pixel at 5500\,\AA\ is in the range 22--46. Radial velocities were measured by cross-correlating the extracted spectra with a G2 numerical mask \citep{Baranne1996}. The uncertainties of the measured RVs are in the range 2.1--8.1\,$\mps$ with a mean value of 4.2\,$\mps$.

\subsubsection{TNG/HARPS-N}
Between July 16$^\mathrm{th}$ 2016 (UT) and May 7$^\mathrm{th}$ 2019 (UT) we collected 14 spectra with the HARPS-N spectrograph \citep[R$\approx$115\,000]{2012SPIE.8446E..1VC} mounted at the 3.58-m Telescopio Nazionale Galileo (TNG) of Roque de los Muchachos Observatory in La Palma, Spain, under the observing programmes A33TAC\_15, OPT17A\_64, CAT17A\_91, and CAT19A\_97. The exposure time was set to 1200--3600, based on weather conditions and scheduling constraints, leading to a SNR per pixel of 15--47 at 5500\,\AA. The spectra were extracted using the off-line version of the HARPS-N DRS pipeline. Doppler measurements and spectral activity indicators were measured using an on-line version of the DRS, the YABI tool\footnote{Available at \url{http://ia2-harps.oats.inaf.it:8000}.}, by cross-correlating the extracted spectra with a G2 mask \citep{Baranne1996}. The uncertainty of the measured RVs are in the range 2.0--8.6\,$\mps$, with a mean value of 4.5\,$\mps$.

\subsubsection{NOT/FIES}
We acquired 6 additional spectra using the FIbre-fed {\'E}chelle Spectrograph \citep[FIES;][]{1999anot.conf...71F,2014AN....335...41T} mounted at the 2.56-m Nordic Optical Telescope (NOT) of Roque de los Muchachos Observatory (La Palma, Spain). The observations were carried out between June 26$^\mathrm{th}$ and September 6$^\mathrm{th}$ 2016 (UT), as part of the OPTICON observing programme 53-109. We used the FIES high-resolution mode, which provides a resolving power of $R$\,=\,67\,000 in the spectral range 3700--7300~{\AA}. Following the observing strategy described in \cite{2010ApJ...720.1118B} and \cite{2015A&A...576A..11G}, we traced the RV drift of the instrument by acquiring long-exposed ThAr spectra ($T_\mathrm{exp}$\,$\approx$\,35\,sec) immediately before and after each science exposure. The exposure time was set to 2700--3600 seconds, according to the sky conditions and scheduling constraints. The data reduction follows standard IRAF and IDL routines, which includes bias subtraction, flat fielding, order tracing and extraction, and wavelength calibration. Radial velocity measurements were computed via multi-order cross-correlations with the RV standard star HD\,50692 \citep{1999ASPC..185..367U}, observed with the same instrument set-up as \host. The SNR per pixel at 5500~\AA\ of the extracted spectra is in the range 15--35. The uncertainties are in the range 6.8--13.6\,$\mps$ with a mean value of 9.7\,$\mps$.

\begin{table}
\caption{HARPS, HARPS-N, and FIES radial velocities (RVs), bisector inverse slope (BIS), and full-width at half maximum (FWHM) of the cross-correlation function.}
\label{table-02}
\begin{tabular}{lrrrrr}
\hline
\hline
BJD$_\mathrm{TDB}$  & RV       & $\sigma_\mathrm{RV}$ & BIS     & $\sigma_\mathrm{BIS}$ & $\mathrm{FWHM}$\\
-$2\,450\,000$      & ($\mps$) & ($\mps$)            & ($\mps$) & ($\mps$)              & ($\kmps$)\\
\hline
\noalign{\smallskip}
\multicolumn{3}{l}{FIES}\\
7566.606106   &   -1236.447   &   10.145  &    -25.604  &    8.240   & 12.240  \\ 
7570.575852   &   -1248.492   &    8.236  &    -19.745  &    5.424   & 12.237  \\
7579.594362   &   -1221.025   &   13.568  &    -15.325  &   10.160   & 12.257  \\
7583.604507   &   -1227.357   &   11.149  &     -1.138  &    9.710   & 12.232  \\ 
7600.515019   &   -1233.204   &    8.028  &    -36.207  &    7.614   & 12.283  \\
7638.387601   &   -1225.100   &    6.816  &    -23.315  &    4.940   & 12.220  \\
\hline
\noalign{\smallskip}
\multicolumn{3}{l}{HARPS-N}\\
7585.595242   &   -1192.158   &   6.794   &   -15.933   &    9.608   &   7.354\\
7603.514891   &   -1191.265   &   5.126   &   -13.175   &    7.249   &   7.363\\
7611.572851   &   -1201.759   &   8.653   &   -15.272   &   12.238   &   7.371\\
7892.699273   &   -1195.302   &   3.475   &    -5.393   &    4.915   &   7.348\\
7921.702566   &   -1191.619   &   3.014   &    -7.642   &    4.263   &   7.359\\
7958.534636   &   -1194.254   &   7.553   &   -39.517   &   10.682   &   7.360\\
7965.540153   &   -1197.765   &   2.357   &   -14.982   &    3.334   &   7.360\\
8013.363503   &   -1186.028   &   1.991   &   -18.747   &    2.816   &   7.366\\
8013.404784   &   -1185.686   &   2.117   &   -30.544   &    2.994   &   7.362\\
8014.362082   &   -1185.942   &   2.722   &   -28.539   &    3.850   &   7.352\\
8014.402379   &   -1189.116   &   2.756   &   -24.429   &    3.898   &   7.365\\
8015.359376   &   -1178.088   &   5.620   &   -11.056   &    7.948   &   7.349\\
8015.402150   &   -1187.664   &   5.605   &   -13.530   &    7.927   &   7.368\\
8610.724187   &   -1193.898   &   4.651   &   -17.515   &    6.577   &   7.359\\
\hline
\noalign{\smallskip}
\multicolumn{3}{l}{HARPS}\\
7508.854151   &   -1195.313   &   3.185   &    -5.317   &    4.504   &   7.447\\
7511.892058   &   -1194.114   &   4.462   &    -2.236   &    6.310   &   7.409\\
7609.625616   &   -1195.440   &   2.732   &    19.011   &    3.865   &   7.415\\
7619.566531   &   -1184.749   &   4.551   &    16.474   &    6.437   &   7.413\\
7637.587335   &   -1181.617   &   6.366   &   -13.161   &    9.004   &   7.400\\
7638.596015   &   -1172.262   &   7.600   &   -21.324   &   10.748   &   7.447\\
7639.571271   &   -1181.987   &   6.990   &   -34.526   &    9.885   &   7.389\\
7645.538147   &   -1194.348   &   4.671   &    11.465   &    6.607   &   7.433\\
7682.520265   &   -1194.190   &   5.191   &    17.641   &    7.342   &   7.420\\
7984.617196   &   -1190.758   &   2.370   &    -5.857   &    3.352   &   7.422\\
7986.590931   &   -1203.107   &   3.223   &    -1.852   &    4.559   &   7.421\\
7987.587443   &   -1197.553   &   2.285   &    13.255   &    3.232   &   7.415\\
7990.553698   &   -1195.977   &   2.729   &    24.528   &    3.859   &   7.418\\
7990.593359   &   -1207.601   &   2.817   &   -20.535   &    3.983   &   7.396\\
7992.576732   &   -1189.811   &   2.093   &    16.287   &    2.950   &   7.419\\
7992.609762   &   -1187.701   &   2.759   &    -9.069   &    3.902   &   7.408\\
8003.594697   &   -1191.639   &   3.620   &    -0.841   &    5.110   &   7.401\\
8235.876739   &   -1189.777   &   8.100   &   -22.636   &   11.455   &   7.409\\
\hline
\end{tabular}
\end{table}

\subsection{High contrast imaging}
\label{subsec-observations-hci}
\begin{figure}
\centerline{\includegraphics[angle=0,width=\linewidth]{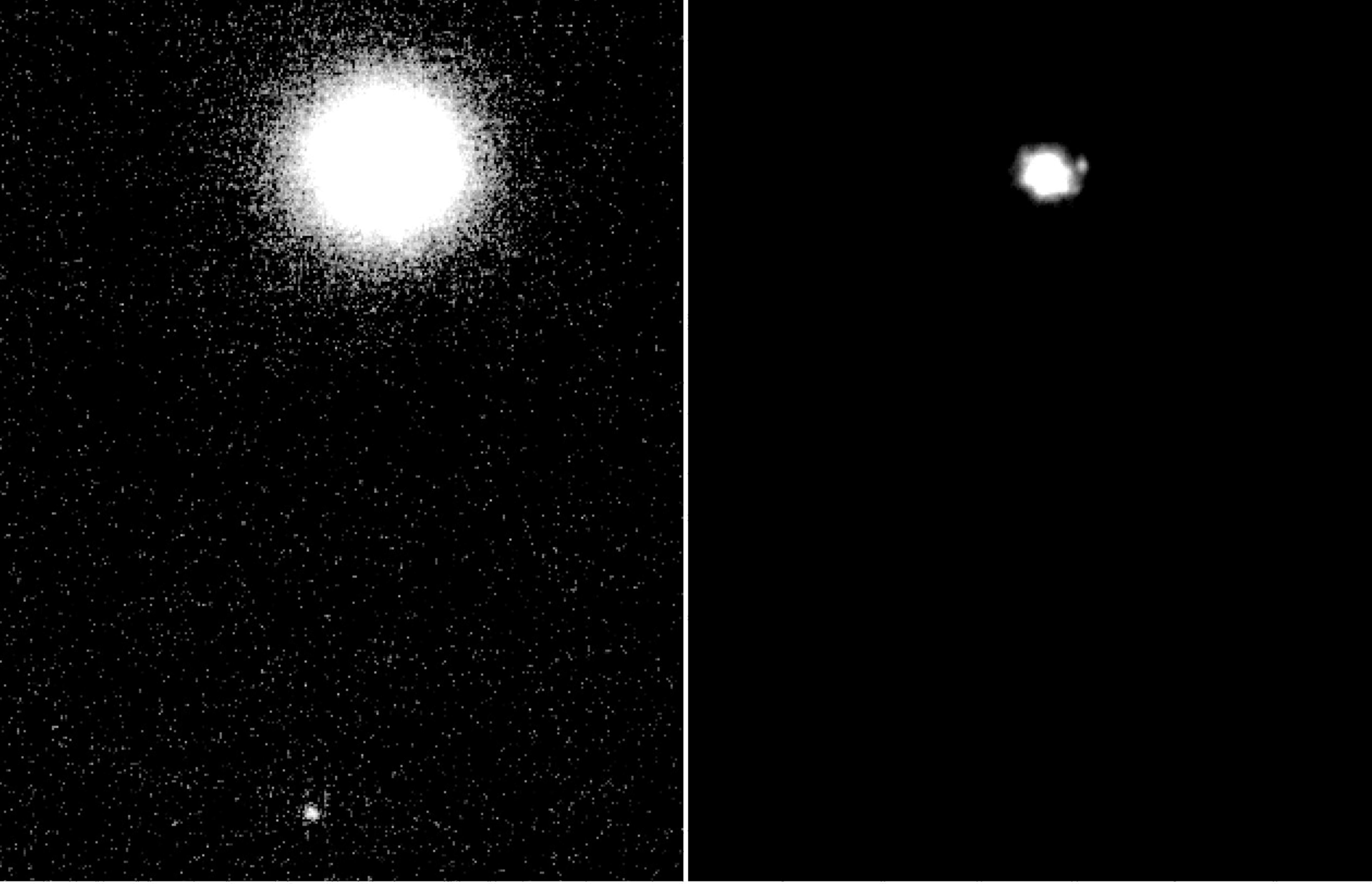}}
\caption{AO image of the surroundings of \host\ obtained with the Gemini-North/NIRI+ALTAIR instrument. Both panels show the same image, with  a FOV of 8.7\arcsec\ in the N-S and 6.7\arcsec\ in the E-W direction (north to the top and east to the left), but with different brightness scales. The left panel shows the star at 6.6\arcsec\ in the S--SE direction and the right one the close neighbor at 0.4\arcsec\ W--NW of \host.
\label{figure-03}}
\end{figure}

To search for nearby stars and estimate a potential contamination factor from such sources we used a high contrast image of \host\ publicly available on the ExoFOP-K2 website\footnote{See \url{https://exofop.ipac.caltech.edu/k2/edit_target.php?id=216494238}.}. The image was acquired on June 19$^\mathrm{th}$ 2016 (UT) using the Frederick C. Gillett Gemini North telescope and its adaptive optics (AO) system facility, ALTAIR with a natural guide star along with a Near InfraRed Imager and spectrograph (NIRI) \citep{2003PASP..115.1388H} using the Brackett Gamma (Br$\gamma$) filter (Gemini-North ID G0218) centered at 2.17~$\microm$, under Gemini Science Program GN-2016A-LP-5. Two faint stars are visible on the Gemini-North/NIRI+ALTAIR AO image of \host\ (Figure~\ref{figure-03}): a very close-in companion at $\sim$0.4\arcsec\ west--north-west (W--NW), and a distant source at $\sim$6.6\arcsec south--south-east (S--SE) of \host. We carefully analyzed the Gemini-North/NIRI+ALTAIR AO image of \host. Table~\ref{table-03} reports separations, position angles, the magnitude difference $\Delta m_{Br\gamma}$, and the $\Delta F_{Br\gamma}$ flux-ratio of these two objects relative to \host. Their brightness ratio at 2.17~$\microm$ is comparable to the observed \ktwo\ transit depth (3500\,ppm), which requires their consideration as sources of false positives (see Section~\ref{subsec-faint_ao_companions}).

\section{Properties of the host star}
\label{sec-host_star}

\subsection{{\bf \emph{Gaia}} measurements}
\host\ is among a small sub-sample of ESA's \gaia\ mission \citep{2016A&A...595A...1G} targets for which the \gaia\ DR2 \citep{2018A&A...616A...1G}\footnote{Released on April 25$^\mathrm{th}$, 2018.} -- a first \gaia-only catalogue -- provides not only astrometric measurements, but also astrophysical parameters (radii, luminosities, extinctions, and reddening) and median radial velocities. \gaia\ DR2 astrometric parameters of \host\ are included in Table\,\ref{table-01}. \gaia\ DR2 values of stellar radius and median RV of \host\ agree with the values determined in the subsections below.

\subsection{Photospheric parameters and stellar rotation velocity measurements using SME}
\label{subsec-sme}
We followed the procedure described in \citet{2017A&A...604A..16F} and \citet{2018A&A...618A..33P} and analysed the co-added spectra from HARPS, HARPS-N, and FIES with the spectral analysis package Spectroscopy Made Easy \citep[{\tt{SME}};][]{1996A&AS..118..595V,2005ApJS..159..141V,2017A&A...597A..16P} to derive the effective temperature $\steff$, surface gravity $\logsg$, iron abundance [Fe/H], and projected rotational velocity $\svrotsini$. Spectroscopy Made Easy uses grids of atmosphere models to calculate synthetic stellar spectra, which are fitted to the observed spectra using a $\chisq$-minimising procedure. We use the line wings of H$_{\alpha}$, which is rather insensitive to $\logsg$ for this spectral type, to model $\steff$ (with a fixed $\logsg$), and the line wings of the Ca\,{\sc i} triplet to model $\logsg$ (with a fixed $\steff$).  We used the latest version of the software (5.2.2) and line lists from the Vienna atomic line database\footnote{{\url{http://vald.astro.uu.se}}.}. The model spectra were taken from {\tt{ATLAS12}} \citep{2013ascl.soft03024K}. The calibration equations for Sun-like stars from \citet{2010MNRAS.405.1907B} and \citet{2014MNRAS.444.3592D} were adopted to fix the micro- and macroturbulent velocities, $\svmic$ and $\svmac$ to 0.5 and 1.0~$\kmps$, respectively. The spectroscopic parameters derived from the HARPS, HARPS-N, and FIES co-added spectra agree well within their nominal error bars. The final adopted values are $\steff$\,=\,5500\,$\pm$\,100\,K, $\logsg$\,=\,4.00\,$\pm$\,0.10\,(cgs), and [Fe/H]\,=\,0.33\,$\pm$\,0.08\,dex (Table~\ref{table-01}). They are defined as the weighted mean of the individual parameters derived from the HARPS, HARPS-N, and FIES co-added spectra.

\subsection{Photospheric parameters and radius measurements using {\tt SpecMatch-emp}}
\label{subsec-specmatch-emp}

As a sanity check, we also analysed the co-added HARPS and HARPS-N spectra using the {\tt SpecMatch-emp} software package \citep{2017ApJ...836...77Y}. {\tt SpecMatch-emp} estimates the stellar effective temperature $\steff$, radius $\sr$, and iron abundance $\sfeh$ by fitting the spectral region between 5000 and 5900\,{\AA} to hundreds of library spectra gathered by the California Planet Search programme. Following the procedure described in \citet{2018AJ....155..127H}, we reformatted the co-added HARPS and HARPS-N spectra so that they can be read by {\tt SpecMatch-emp}. We found $\steff$\,=\,5597\,$\pm\,$110\,K and $\sfeh$\,=\,0.33\,$\pm$\,0.08 dex, which agree with the effective temperature and iron abundance determined with SME (Table~\ref{table-01}) within 1$\sigma$. We found also that \host\ is a slightly evolved star with a stellar radius of $\sr$\,=\,1.33\,$\pm$\,0.21\,$\Rea$. We finally obtained a first estimate of the stellar mass ($\sm$\,=\,1.16\,$\pm$\,0.08$~\Mea$) via Monte Carlo simulations using the empirical equations by \citet{2010A&ARv..18...67T} alongside $\steff$, $\sfeh$, and $\sr$.

\begin{table}
\caption{Relative properties of the two nearby stars to  \host\ detected  with
the Gemini-North/NIRI+ALTAIR.
\label{table-03}}
\begin{center}
\begin{tabular}{lcc}
\hline
\hline
\noalign{\smallskip}
Parameter                      & W--NW              & S--SE \\
                               & close-in star      &  distant star\\

\noalign{\smallskip}
\hline
\noalign{\smallskip}
Separation (\arcsec)           & $  0.38 \pm 0.011$                 & $6.598 \pm 0.011$\\
Position Angle (deg)           & $286.2  \pm 1.5  $                 & $173.3 \pm 1.5$\\
$\Delta m_{Br\gamma}$ (mag)    & $  4.72 \pm 0.15 $                 & $ 6.65 \pm 0.15$\\
$\Delta F_{Br\gamma}$ relative flux   & $  (1.3 \pm 0.2) \times 10^{-2}$   & $ (2.2 \pm 0.4) \times 10^{-3}$\\
\noalign{\smallskip}
\hline
\end{tabular}
\end{center}
\end{table}

\subsection{Physical parameters}
\label{subsec-param}

We refined the fundamental parameters of \host\ utilising the web interface\footnote{Available at \url{http://stev.oapd.inaf.it/cgi-bin/param_1.3}.} \texttt{PARAM~1.3} along with \texttt{PARSEC} isochrones  \citep{2012MNRAS.427..127B}. Following the method described in \citet{2008ApJ...687.1303G}, we found that the interstellar extinction along the line of sight to the star is $A_\mathrm{v}\,=\,0.10\,\pm\,0.05$. Using the effective temperature and iron abundance derived in Section~\ref{subsec-sme}, alongside the extinction-corrected visual magnitude and the \gaia\ parallax\footnote{We accounted for \gaia\ systematic uncertainties adding quadratically 0.1\,mas to the nominal uncertainty of parallax \citep{2018A&A...616A...9L}.} (Table~\ref{table-01}), we determined a mass of $\sm$\,=\,1.03\,$\pm$\,0.03\,$\Msun$ and a radius of $\sr$\,=\,1.28\,$\pm$\,0.07\,$\Rsun$, which agree with the values derived in Section~\ref{subsec-specmatch-emp}. Stellar mass and radius implies a surface gravity of $\logsg$\,=\,4.21\,$\pm$\,0.05 (cgs), which is higher than our spectroscopic value of $4.0\,\pm\,0.1$ (cgs), but within its 2$\sigma$ error bars. The age of the star was constrained to be 8.9\,$\pm$\,1.7\,Gyr, further confirming the evolved status of \host. The values of stellar radius and mass agree within 3-$\sigma$ with the ones determined by \citet{2018AJ....155...21P} ($\sr = 1.45_{-0.18}^{+0.20}$\,$\Rsun$, $\sm = 1.17_{-0.08}^{+0.10}$\,$\Msun$), \citet{2018AJ....155..136M} ($\sr = 1.064_{-0.047}^{+0.069}$\,$\Rsun$, $\sm = 1.101_{-0.028}^{+0.025}$\,$\Msun$), and \citet{2018AJ....156..277L} ($\sr = 1.28 \pm 0.03$\,$\Rsun$, $\sm = 1.11 \pm 0.04$\,$\Msun$). We stress that the parameter estimates determined in the three works listed above are based on spectra with relatively low SNR, in contrast to our co-added, high SNR, HARPS, HARPS-N and FIES spectra. \citet{2018AJ....155...21P} and \citet{2018AJ....156..277L} used the same spectra collected with the HIgh Resolution Echelle Spectrometer \citep[HIRES][]{1994SPIE.2198..362V} mounted on 10-m Keck I telescope, with typical SNR\,=\,45 for stars with $V < 13.0$. \citet{2018AJ....155..136M} used spectra collected with Tillinghast Reflector Echelle Spectrograph (TRES) mounted on the 1.5-m Tillinghast telescope at the Whipple Observatory on Mt. Hopkins in Arizona with even lower SNR.

We also calculated the $UVW$ space velocities of \host\ using the IDL code {\tt gal\_uvw}\footnote{Available at \url{https://idlastro.gsfc.nasa.gov/ftp/pro/astro/gal_uvw.pro}.} \citep[based upon][]{JohnsonSoderblom87}, using the \gaia\ DR2 proper motions and parallax, and the average of the HARPS and HARPS-N systemic velocities $\gamma$ (Table~\ref{table-01}). Our calculated values of $UVW$ are listed in Table~\ref{table-01}; we quote values in the local standard of rest using the solar motion of \cite{Coskunoglu11}. We then used the methodology of \cite{Reddy06} to determine the Galactic population membership of \host. We found that \host\ has a $>99\%$ probability of belonging to the Galactic thin disk, and less than $1\%$ of belonging to either the thick disk or the halo. This is consistent with \host's high metallicity of $\sfeh$ = \sfehv.

The final adopted stellar parameters are listed in Table~\ref{table-01}. The effective temperature and surface gravity translate into a G7\,V spectral type \citep{2009ssc..book.....G}.

\subsection{Faint AO companions}
\label{subsec-faint_ao_companions}
From the two faint companions to \host\ identified in the Gemini-North/NIRI+ALTAIR AO image (Section~\ref{subsec-observations-hci}), the one located 6.6\arcsec\ S--SE of \host\ was identified in the \gaia\ DR2 as the source 6772454206445987712. Based on its very small proper motion ($\mathrm{PM_{RA}}$\,=\,0.29\,$\pm$\,0.52$\,\maspyr$ and $\mathrm{PM_{DEC}}$\,=\,$-$0.92\,$\pm$\,0.45$\,\maspyr$) and distance found by \citet{2018AJ....156...58B} ($d = 5.103^{+3.435}_{-2.094}$ kpc), we concluded that it is a background star. Using the \gaia\ G-band magnitude ($G\,=\,18.765\pm0.010$), we derived a G-band brightness ratio relative to \host\ of 0.0027\,$\pm$\,0.0001. Considering the close similarity between the \gaia\ G-band and the \kepler\ passband, this companion is too faint to be the source of the transit signal detected in the \ktwo\ data.

For the close-in W--NW companion we cannot determine whether it is physically bound or unbound to \host. Yet, its very small angular separation of 0.4\arcsec\ supports the binary scenario for \host. Based on the \texttt{Besan\c{c}on} Galactic population model\footnote{Available at \url{http://modele2016.obs-besancon.fr}.} \citep{2003A&A...409..523R} and following the procedure described in \cite{2019MNRAS.484.3522H} we calculated the probability of a chance alignment to be $0.04\%$. Assuming that the W--NW companion is physically bound to \host, we can then obtain further information about it.

The central wavelength of 2.19~$\microm$ of the Br$\gamma$ filter is nearly identical to that of the near-infrared $K$ band. Therefore we used the apparent $K$ magnitude of \host\ from Table~\ref{table-01} ($m_K=10.765 \pm 0.019$) and the magnitude difference from Table~\ref{table-03} to calculate absolute $K$ magnitudes of both stars. They are equal to $M_K=2.778 \pm 0.090$ for \host\ and $M_K=7.50 \pm 0.22$ for the nearby companion. Making use of the Dartmouth isochrone table \citep{2008ApJS..178...89D} for metallicity $\sfeh = 0.36$ and ages between 9 and 11 Gyr, we estimated that the nearby companion is a M3.5--M4  red-dwarf with a mass between 0.21 and 0.28\,$\Msun$. Using its angular separation from Table~\ref{table-03} and the DR2 parallax of \host\, we calculated a lateral separation from \host\ of $150.4^{+8.2}_{-7.7}$\,au. We note that current models of planetary formations in wide binary stellar systems predict a shortage of giant planets in binaries with separations of $\leq$100\,au \citep[e.g.,][]{2000ApJ...537L..65N,2005MNRAS.363..641M,2006Icar..183..193T}; the nearby companion should therefore not have affected the formation of the \host\ planetary system.

Based on the Dartmouth isochrone table for metallicity $\sfeh = 0.36$ and ages between 9 and 11 Gyr, we also estimated the nearby star's absolute \kepler\ magnitude ($M_{K_{p}}$) as 10.75--11.5\,mag, and its apparent \kepler\ magnitude as 18.75--19.5\, mag. That is, its \kepler\ brightness is 0.0019 $\pm$ 50\% of \host's brightness. However, a false-positive scenario with an equal mass eclipsing binary (eclipse depth equal to 50\%) and a transit signal with a depth of $3.5 \times 10^{-3}$ can only be caused by a binary that is brighter than 0.007 times the host's brightness. Therefore, assuming that the nearby W--NW star is physically bound with \host, we may exclude it as a source of a false positive.

\section{Global Analysis}
\label{sec-global_analysis}

We used the code \href{https://github.com/oscaribv/pyaneti}{\texttt{pyaneti}} \citep{2019MNRAS.482.1017B} to perform the joint analysis of the RV and \ktwo\ transit data. The code uses the limb-darkened quadratic model by \citet{2002ApJ...580L.171M} to fit the transit light curves and a Keplerian model for the RV measurements. We integrated the light curve model over 10 steps to simulate the \kepler\ long-cadence integration \citep{2010MNRAS.408.1758K}. Fitted parameters, parametrizations and likelihood are similar to previous analysis performed with \texttt{pyaneti} \citep[e.g.][]{2016AJ....152..193B,2018MNRAS.475.1765B}.

The photometric data includes $\sim$17 hours (i.e, twice the transit duration) of data-points centered around each of the 4 transits observed by \ktwo. We de-trended the photometric chunks using the program \href{https://github.com/oscaribv/exotrending}{\texttt{exotrending}}. \citep{2017ascl.soft06001B}, fitting a second order polynomial to the out-of-transit data. The Doppler measurements include the 6 FIES, 14 HARPS-N, and 18 HARPS RVs presented in Section~\ref{subsec-observations-hrs}.

We adopted uniform priors for all the parameters; details are given in Table~\ref{table-04}. We started 500 Markov chains randomly distributed inside the prior ranges. Once all chains converged\footnote{We define convergence as when chains have a scaled potential factor < 1.02 for all the parameters \citep[see ][for more details]{Gelman1992}.}, we ran 5000 additional iterations. We used a thin factor of 10 to generate a posterior distribution of 250,000 independent points for each parameter.

We first explored the properties of the Doppler signal by fitting the RV data alone. We tested different models: one model assumes there is no Doppler reflex motion; one model assumes the presence of a planet on a circular orbit; another model assumes the presence of a planet on an eccentric orbit. These three models were run with and without a jitter term for each instrument. This generates a set of 6 different models. The main statistical properties of each model are listed in Table~\ref{tab:models}. From this Table we can draw the following conclusions: 1) the models including a planet signal are strongly preferred over the models without it; 2) the eccentric model is preferred, as suggested also by the transit fit (see the following paragraph); 3) the model does not require to add a jitter term for each spectrograph, suggesting that any extra signal (stellar variability, other planets, etc.) are below the instrumental precision. This supports our RV analysis assuming only a Keplerian orbit. We note that we still fit for a jitter term for each instrument to allow more flexibility to our modelling and to mitigate the effects of the relatively sparse sampling of our data on the accuracy of the semi-amplitude estimate.

\begin{table*}
\begin{center}
\caption{Model comparison for our RV fits.\label{tab:models}}
\begin{tabular}{lcccccr}
\hline
\hline
\noalign{\smallskip}
 Test&    Npars   &     log likelihood  &  BIC &                 K(m/s) \\
 \hline
No planet - no jitter     &                                     3          &           118     &         -226          &         0  \\
No planet - jitter  &                                              6          &           135   &          -248       &             0  \\
planet - circular orbit- no jitter    &                   6        &             136  &           -249          &         $7.20 \pm 1.15$ \\
planet - circular orbit - jitter  &                         9            &         143     &         -254            &      $ 7.18 \pm 1.60$ \\
planet - eccentric orbit - no jitter  &                8               &       154      &       -280          &        $ 9.31 \pm 1.20 $ \\
planet - eccentric orbit - jitter  &                     11        &            154       &       -269            &     $ 9.27 \pm 1.30 $ \\
\hline
\end{tabular}
  \begin{tablenotes}\footnotesize 
  \item \emph{Note} -- Further details about the Bayesian Information Criterion (BIC) are given in, e.g., \citet[][]{BurnhamAnderson2002}.
\end{tablenotes}
\end{center}
\end{table*}

We used Kepler's third law to check if the stellar density derived from the modeling of the transit light curves is consistent with an eccentric orbit \citep[see, e.g.,][]{2015ApJ...808..126V}. We first ran an MCMC analysis assuming the orbit is circular. The derived stellar density is $0.32^{+0.02}_{-0.06}\,{\rm g\,cm^{-3}}$. This density disagrees with the stellar density of \densspb\ obtained from the spectroscopic parameters derived in Section~\ref{sec-host_star}. We then performed a joint analysis allowing for an eccentric solution. We derived a stellar density of $0.82_{-0.35}^{+0.38}\,{\rm g\,cm^{-3}}$, which is consistent with the spectroscopically derived stellar density.
This provides further evidence that the planetary orbit is eccentric. For the final analysis, we decided to set a Gaussian prior on $a/\sr$ using Kepler's third law and the stellar mass and radius derived in Section~\ref{sec-host_star} and listed in Table~\ref{table-01}.

\begin{figure}
\includegraphics[width=0.475\textwidth]{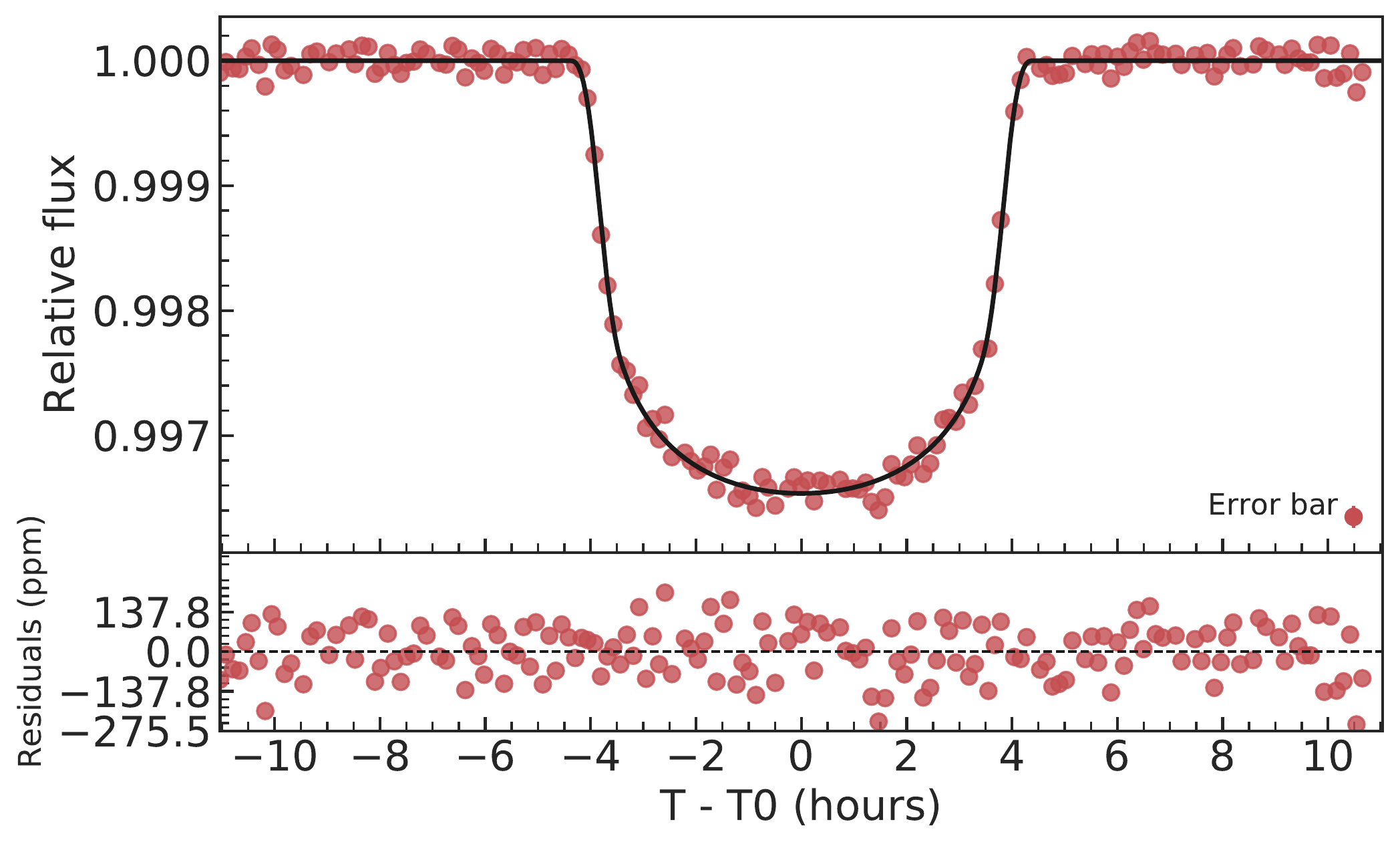} 
\includegraphics[width=0.475\textwidth]{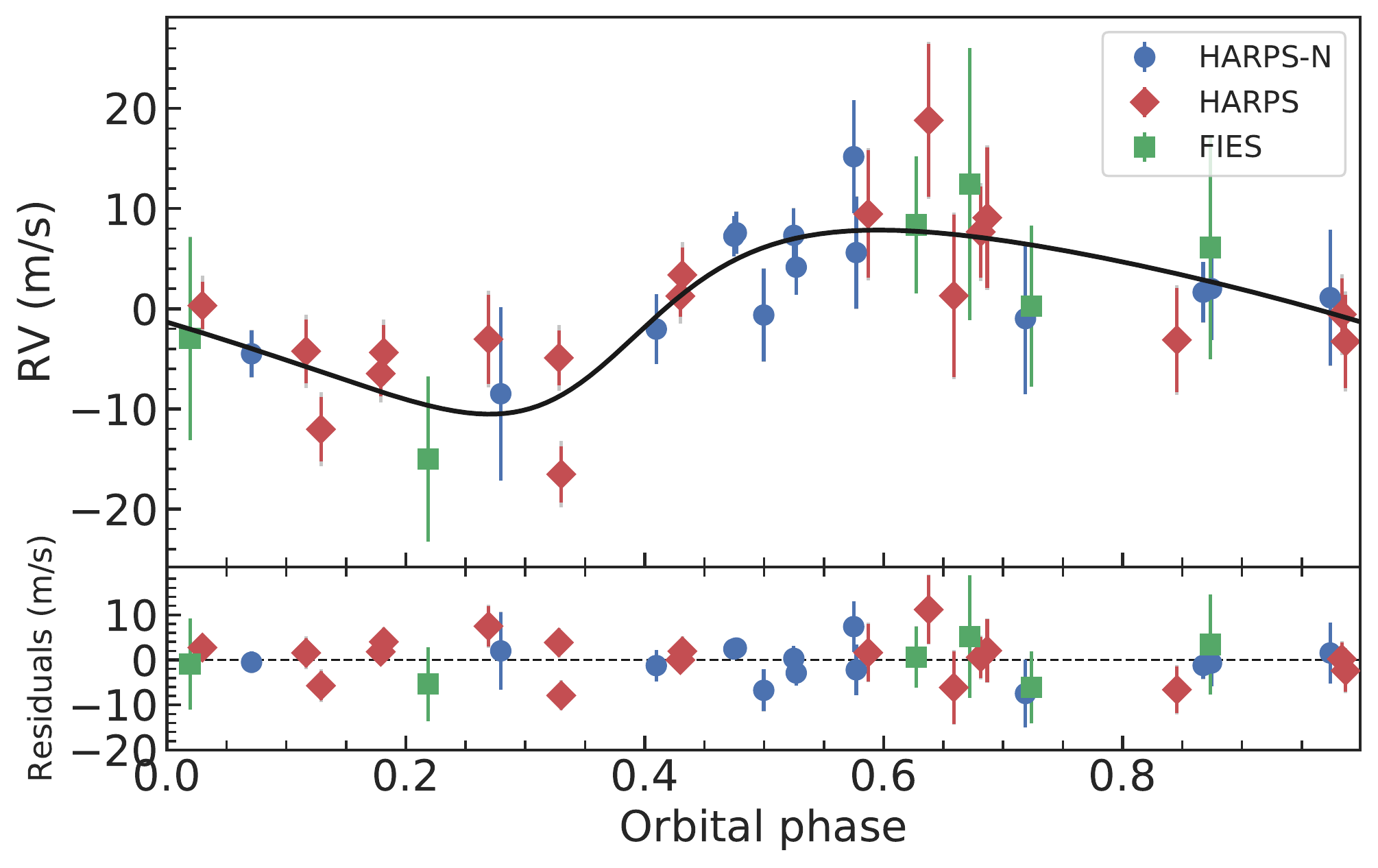}
\caption{\emph{Top panel:} Transit light curve folded to the orbital period of \planet\ and residuals. The thick black line is the re-binned best-fitting transit model. The red points are the \ktwo\ data. \emph{Bottom panel:} The RV curve of \planet\ phase-folded to the orbital period of the planet. The best fitting solution is marked with a solid black line. HARPS-N, HARPS, and FIES data are shown with blue circles, red diamonds, and green squares, respectively. The lower panel shows the residuals to the best-fitting model.
\label{figure-04}}
\end{figure}

The median and 68.3\% percent credible intervals of the marginalized posterior distributions are reported in Table \ref{table-04}. Figure~\ref{figure-04} displays the RV and transit data together with the best fitting model. We show a corner plot of the fitted parameters in Figure~\ref{figure-09}.

\begin{figure}
\includegraphics[width=\linewidth]{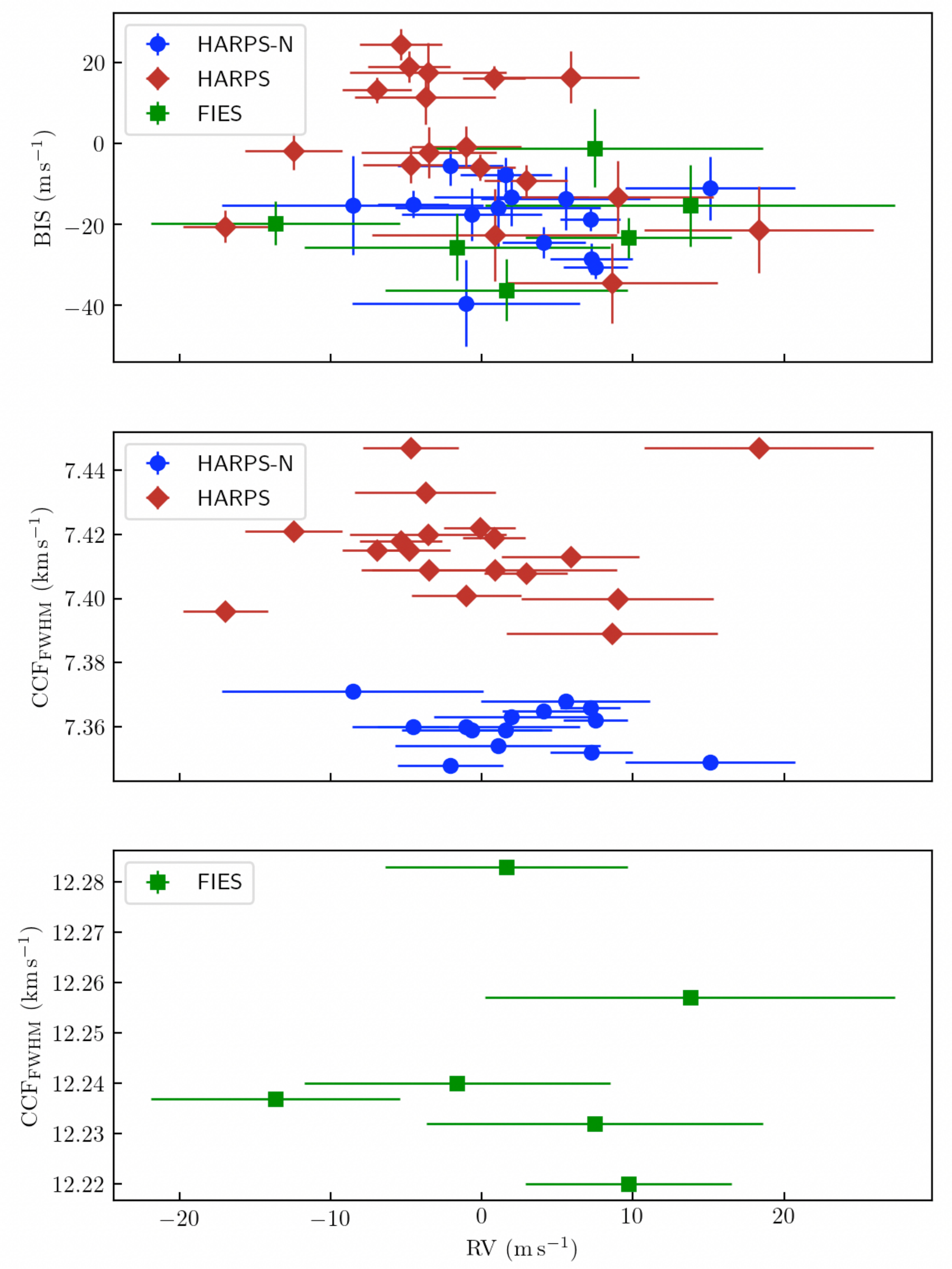}
\caption{\emph{Top panel:} The BIS versus RVs from HARPS-N, HARPS, and FIES. \emph{Middle panel:} The $\mathrm{CCF_{FWHM}}$ versus RVs from HARPS-N and HARPS. \emph{Bottom panel:} The $\mathrm{CCF_{FWHM}}$ versus RVs from FIES. RVs from all instruments have been subtracted by the systemic velocities listed in Table~\ref{table-04} and derived by our joint analysis.
\label{figure-05}}
\end{figure}

The HARPS, HARPS-N, and FIES Doppler measurements show an RV variation in phase with the transit ephemeris (Figure~\ref{figure-04}, lower panel). However, as described by \cite{2013A&A...550A..75C}, contaminant stars that are within the sky-projected angular size of the spectrograph fibre (1\arcsec\ for HARPS and HARPS-N, 1.3\arcsec\ for FIES) may affect the radial velocity measurements of the target star. If the radial velocity of the contaminant star is changing, i.e., its spectrum is shifting across the spectrum of target star, it can distort the spectral line profile of the target (and hence its CCF), mimicking the presence of an orbiting planet. As presented by \cite{2013A&A...550A..75C} in their Table~8, for magnitude differences of $\sim$5-6\,mag, the impact of F2\,V--K5\,V contaminant star on a G8\,V target star can be as high as 10\,$\mps$. If the nearby N--NW star, which has an angular separation $0.38 \pm 0.011$\arcsec\ from \host, is an F or G background eclipsing binary, it may not only generate a transit-like signal in the light curve of \host\ every 19.9 days, but also a low-amplitude radial velocity signal at this period. We carefully checked the the FWHM and BIS of the HARPS, HARPS-N, and FIES cross-correlation functions to search for potential line profile variation induced by the blend companion. The generalized Lomb-Scargle periodograms \citep{2009A&A...496..577Z} of these indicators show no significant signal neaither at the 19.9-day period and its harmonics, nor at any other period. We also found no correlation between the FWHM and BIS, and the RV measurements (Figure~\ref{figure-05}). In particular, the Spearman correlation coefficient between the HARPS RV measurements and the BIS of CCFs is equal to $r_{RV-BIS,HARPS}$ = -0.45 and between HARPS RVs and FWHM is equal to $r_{RV-FWHM,HARPS}$ = -0.18. The Spearman correlation coefficient between the HARPS-N RVs and BIS is equal to $r_{RV-BIS,HARPS-N}$ = -0.19 and between the HARPS-N RVs and FWHM is equal to $r_{RV-FWHM,HARPS-N}$ = -0.06. In the case of FIES measurements, the Spearman correlation coefficient between RVs and BIS is equal to $r_{RV-BIS,FIES}$ = 0.37 and between RVs and FWHM is equal to $r_{RV-FWHM,FIES}$ = -0.08.

We note that we could not measure the stellar rotation period from the \ktwo\ light curve. Using the stellar radius determined in Section~\ref{subsec-param} and the projected rotation velocity determined in Section~\ref{subsec-sme}, we found the upper limit of the stellar rotation period to be $\sprot = 21.6^{+12.6}_{-6.3}$ days. This means that the stellar rotation period of K2-280 is shorter than 34.2 days. Following the prescription given by \citet{2012MNRAS.419.3147A}, the photometric variation found in the \ktwo{} light curve ($\sim$600\,ppm) implies an activity induced RV signal of about 2\,$\mps$ (1.9\,$\mps$ for stellar rotation period equal to orbital period of \planet{} (19.9 days) or 1.1\,$\mps$ for $\sprot$ equal to 34.2 days). The probability that stellar rotation modulation may generate RV variations of \host{} is therefore very low. We conclude that most likely the Doppler shift of \host{} is induced by the orbital motion of a planet transiting \host{} rather than a blended eclipsing binary or stellar rotation modulation. We stress however that the activity-induced RV signal at a level of $\sim$2\,$\mps$ is larger than the jitter terms listed in Table~\ref{table-04} and larger than the precision on our estimate of the Doppler semi-amplitude variation induced by the planet ($K$\,=\,\kb). Therefore, we warn the reader that our semi-amplitude estimate might be affected by unaccounted for stellar activity.

\begin{table*}
\begin{center}
\caption{\host\ Stellar and Planetary Parameters
\label{table-04}}
  \begin{tabular}{lcc}
  \hline
  Parameter & Prior$^{(\mathrm{a})}$ & Inferred value$^{(\mathrm{b})}$ \\
  \hline
  \multicolumn{3}{l}{\emph{Model Parameters}} \\
  \noalign{\smallskip}
    Orbital period $P_{\mathrm{orb}}$ (days) &  $\mathcal{U}[ 19.89 , 19.90 ]$ & \Pb[] \\
    Transit epoch $T_0$ (BJD$_\mathrm{TDB}-$2\,450\,000) & $\mathcal{U}[ 7307.55 , 7307.65]$ & \Tzerob[]  \\  
    Scaled semi-major axis $a/R_{\star}$ &  $\mathcal{N}[25.58, 0.90]$ & \arb \\
    Scaled planet radius $R_\mathrm{p}/R_{\star}$ & $\mathcal{U}[0,0.2]$ & \rrb  \\
    Impact parameter, $b$  & $\mathcal{U}[0,1]$  & \bb \\
    $\sqrt{e} \sin \omega_\star$ &  $\mathcal{U}[-1,1]$ & \esinb \\
    $\sqrt{e} \cos \omega_\star$  &  $\mathcal{U}[-1,1]$ & \ecosb \\
    Radial velocity semi-amplitude variation $K$ (m s$^{-1}$) & $\mathcal{U}[0,50]$ & \kb[] \\
    Parameterized limb-darkening coefficient $q_1$  & $\mathcal{U}[0,1]$ & \qone \\
    Parameterized limb-darkening coefficient $q_2$  & $\mathcal{U}[0,1]$ & \qtwo \\   
    Systemic velocity $\gamma_{\mathrm{HARPS-N}}$  (km s$^{-1}$) & $\mathcal{U}[  -2.2 , -0.2]$ & \HARPSN[] \\
    Systemic velocity $\gamma_{\mathrm{HARPS}}$  (km s$^{-1}$) & $\mathcal{U}[ -2.2 , -0.2]$ & \HARPS[] \\
    Systemic velocity $\gamma_{\mathrm{FIES}}$  (km s$^{-1}$) & $\mathcal{U}[ -2.2 , -0.2]$ & \FIES[] \\
    Jitter term $\sigma_{\rm HARPS-N}$ (m s$^{-1}$) & $\mathcal{U}[0,100]$ & \jHARPSN[] \\
    Jitter term $\sigma_{\rm HARPS}$ (m s$^{-1}$) & $\mathcal{U}[0,100]$ & \jHARPS[] \\
    Jitter term $\sigma_{\rm FIES}$ (m s$^{-1}$) & $\mathcal{U}[0,100]$ & \jFIES[] \\
    \hline
    \multicolumn{3}{l}{\emph{Derived Parameters planet b}} \\
    Planet mass $M_\mathrm{p}$ ($M_{\rm \oplus}$) & $\cdots$ & \mpb[]  \\
    Planet radius $R_\mathrm{p}$ ($R_{\rm \oplus}$) & $\cdots$ & \rpb[] \\
    Planet density $\ppden$ (${\rm g\,cm^{-3}}$) & $\cdots$ & \denpb[] \\
    Semi-major axis of the planetary orbit $a$ (au) & $\cdots$ & \ab[]  \\
    Orbital eccentricity, $e$ &  $\cdots$ & \eb \\
    Angle of Periastron, $\omega_\star$ (deg) &  $\cdots$ & \wb[] \\
    Time of periastron $T_{\rm p}$ (BJD$_\mathrm{TDB}-$2\,450\,000) & $\cdots$ & \Tperib[] \\
    Transit duration $\tau_{14}$ (hours) & $\cdots$ & \ttotb[] \\
    Equilibrium temperature$^{(\mathrm{c})}$  $T_\mathrm{eq}$ (K)  & $\cdots$ &  \Teqb[] \\
    Linear limb-darkening coefficient $u_1$ & $\cdots$ & \uone \\
    Quadratic limb-darkening coefficient $u_2$ & $\cdots$ & \utwo\\
    Planet surface gravity$^{(\mathrm{d})}$ (${\rm cm\,s^{-2}}$) & $\cdots$ & \grapb[] \\
    Planet surface gravity (${\rm cm\,s^{-2}}$)  & $\cdots$ & \grapparsb[] \\
   \noalign{\smallskip}
  \hline
  \end{tabular}
\end{center}
\emph{Note} -- $^{(\mathrm{a})}$ $\mathcal{U}[a,b]$ refers to uniform priors between $a$ and $b$, and $\mathcal{N}[a,b]$ to Gaussian priors with median $a$ and standard deviation $b$.  $^{(\mathrm{b})}$  The inferred parameter value and its uncertainty are defined as the median and 68.3 percent credible interval of the posterior distribution.  $^{(\mathrm{c})}$ Assuming albedo $= 0$. $^{(\mathrm{d})}$ Calculated from the scaled-parameters as suggested by \citet[][]{Sotuhworth2007}. 
\end{table*}

\section{Discussion and Summary}
\label{sec-disasum}
\subsection{\planet\ and the current sample of sub-Saturn planets}

\begin{figure}
\includegraphics[width=\linewidth]{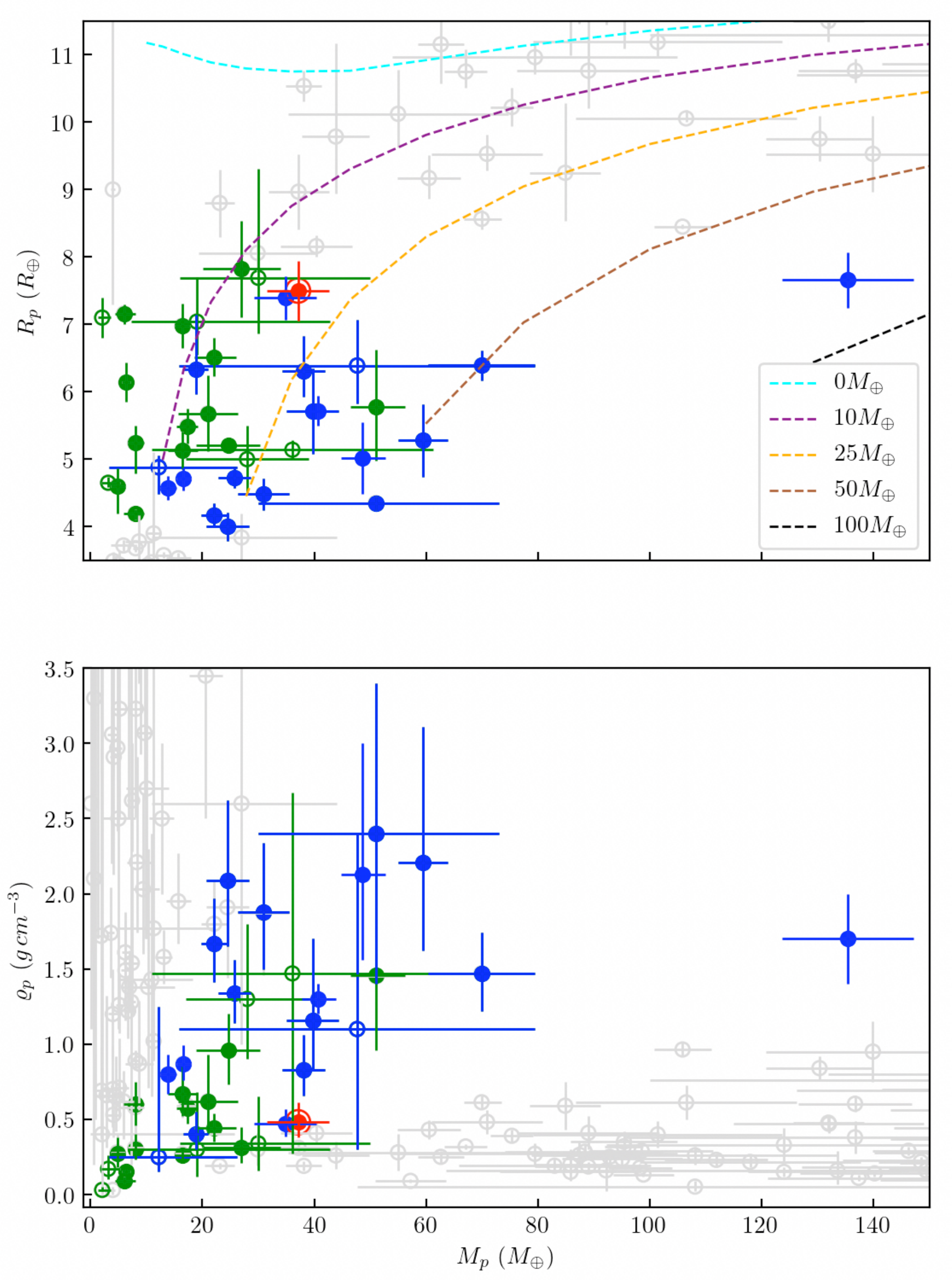}
\caption{Mass--radius (upper panel) and mass--density (lower panel) diagrams for a sample of sub-Saturns ($R_\mathrm{p}\,= 4-8\,\Rea$). Sub-Saturns whose mean densities have been measured with a precision better than 50\% located in multi-planet systems are marked with green filled circles, wheres those in single systems are marked with blue filled circles. The position of \planet\ is indicated as a red-rimmed circle. Sub-Saturns with densities measured with a precision slightly worse than 50\% are marked with green and blue open circles. The remaining planets with measured radii, masses and mean densities (NASA Exoplanet Archive \citep[][]{2013PASP..125..989A}, as of July 2019) are marked with open gray circles. The dashed lines on the mass--radius diagram (upper panel) correspond to the \citet{2007ApJ...659.1661F} models for planet core masses of 0, 10, 25, 50, and 100\,$\Mea$ and age 10\,Gyrs.
\label{figure-06}}
\end{figure}

With a mass of $\pbm$\,=\,\mpb\ and a radius of $\pbr$\,=\,\rpb, \planet\ joins the group of sub-Saturns planets -- defined as planets having radii between 4 and 8~$\Rea$ \citep{2017AJ....153..142P} -- whose masses and radii have been measured. The basic physical parameters of a sample of 23 sub-Saturns with densities measured with a precision better than 50\% have been presented and discussed by \cite{2017AJ....153..142P}. We here extend this sample by adding \planet\ alongside 6 additional sub-Saturns that have densities measured with a precision better than 50\%, as described below. WASP-156\,b \citep{2018A&A...610A..63D}, a $\sim$0.5\,$\Rjup$ planet with a Jupiter-like density was discovered by the ground-based SuperWASP transit survey \citep{2006PASP..118.1407P,2014CoSka..43..500S}. Kepler-1656\,b, a dense sub-Saturn with a high eccentricity of $e$\,=\,0.84 transiting a relatively bright ($V = $\,11.6\,mag) solar-type star, was recently reported by \cite{2018AJ....156..147B}. Three sub-Saturns were discovered and characterised by the KESPRINT consortium, two of them in \ktwo\ campaign 3 \citep[K2-60\,b,][]{2017AJ....153..130E} and campaign 14 \citep[HD\,89345\,b, aka K2-234\,b,][]{2018MNRAS.478.4866V,2018AJ....156..127Y}, and HD\,219666\,b \citep{2019A&A...623A.165E} in \tess\ Sector~1. One sub-Saturn, GJ\,3470\,b \citep{2012A&A...546A..27B}, orbiting an M1.5~dwarf was not included by \cite{2017AJ....153..142P}, but we add it to the current sample, adopting the parameters from \cite{2016MNRAS.463.2574A}.  All of these new sub-Saturns, including \planet, reside in apparently single systems. Figure~\ref{figure-06} shows the mass--radius and mass--density diagrams for this extended sample of 30 planets. Sub-Saturns found to be in multi-planet systems are marked with green filled circles, whereas those in single systems are marked with blue filled circles. The position of \planet\ is indicated with a red-rimmed circle. Sub-Saturns whose density has been measured with a precision slightly worse than 50\% are marked with green and blue open circles. All the remaining transiting planets with measured radii and masses are marked with open gray circles\footnote{As retrieved from the NASA Exoplanet Archive \citep[][]{2013PASP..125..989A} -- July 2019.}. According to the \citet{2007ApJ...659.1661F}'s models -- also shown in the mass-radius diagram (Figure~\ref{figure-06}, upper panel) -- \planet\ has a core of about 10--25~$\Mea$, accounting for $\sim$25--65\% of its total mass.

The diagrams in Figure~\ref{figure-06} confirm the main characteristics found by \cite{2017AJ....153..142P} for the population of sub-Saturns. One of the main property is the uniform distribution of masses in the range $\sim$5--75~$\Mea$. With a mass of 135\,$\pm$\,12\,$\Mea$ and radius of 7.66\,$\pm$\,0.41\,$\Rea$,  K2-60\,b \citep{2017AJ....153..130E} is close to the lower envelope of giant planets on the mass--radius diagram and is the only sub-Saturn-sized planet with a mass higher than Saturn (95.16\,$\Mea$). With a mean density of 1.7\,$\pm$\,0.3~$\gpcmcmcm$ (i.e. Neptune's density), K2-60\,b is also the most dense planet in the mass range $\sim$75--250~$\Mea$. As stressed by \cite{2017AJ....153..130E}, K2-60\,b with radius smaller than expected from the models of \citet{2011ApJ...729L...7L} is more dense than expected and close to the sub-Jovian desert characterised by scarcity of planets with orbital periods below 4 days and masses lower than $\sim$300\,$\Rea$ \citep{2011ApJ...727L..44S,2013ApJ...763...12B,2016A&A...589A..75M}. The underestimation of its radius was excluded based on adaptive optics imaging \citep{2016AJ....151..159S}. Only radial accelerations lower than 2\,$\mpspd$ that can not be excluded based on RVs collected by \cite{2017AJ....153..130E} suggest that mass of K2-60\,b may be lower than current determination. Nevertheless, this intriguing planet may help with a study of sub-Jovian desert and its borders.

\begin{figure}
\includegraphics[width=\linewidth]{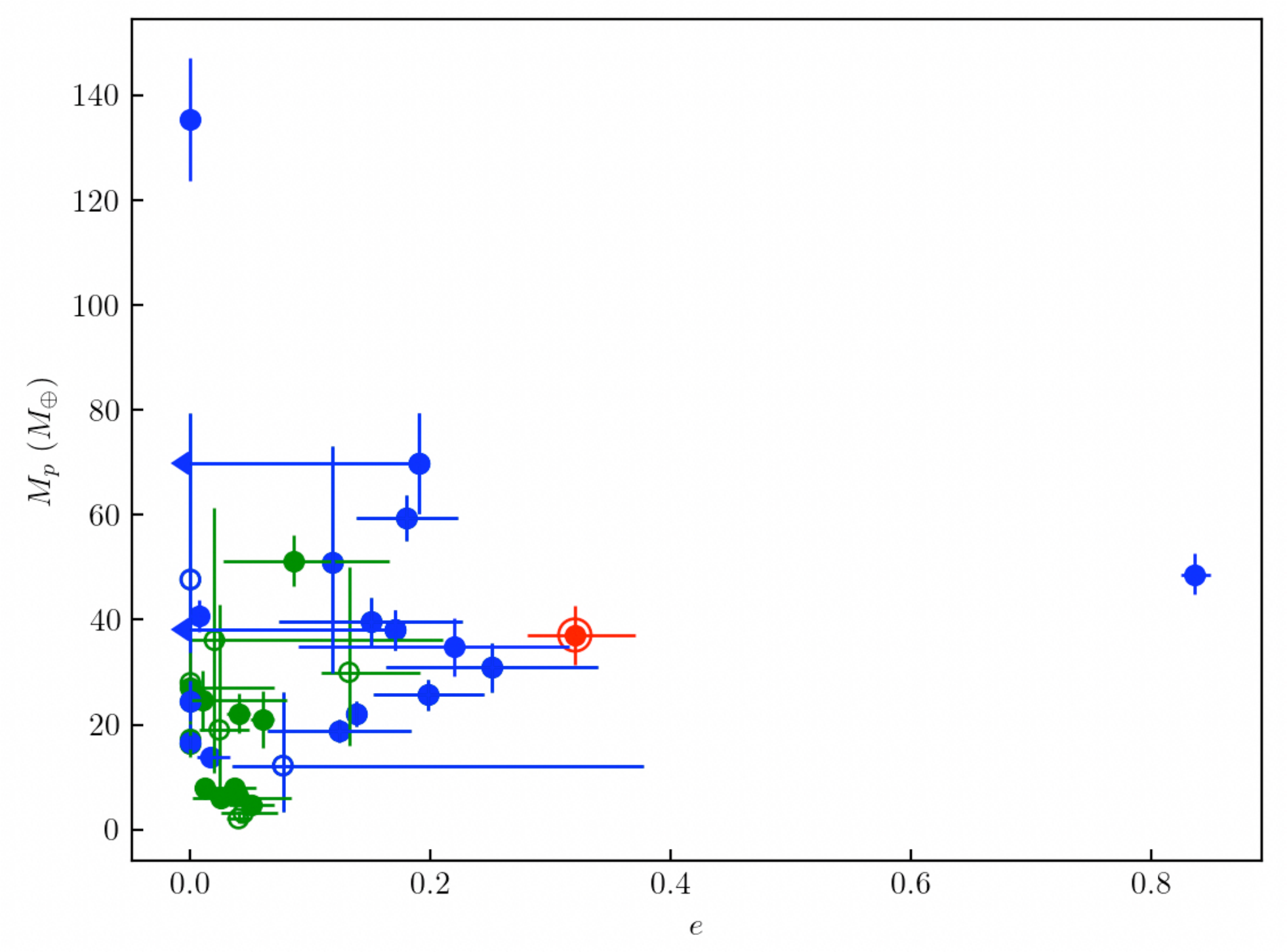}
\caption{Mass of sub-Saturn planets as a function of the eccentricity. Samples and point symbols as in Figure~\ref{figure-06}.
\label{figure-07}}
\end{figure}

Although the mass distribution of sub-Saturns is quite uniform, the most massive ones have radii close to and below $\sim$6~$\Rea$, visible as a correlation on the mass--density diagram (Figure~\ref{figure-06}, lower panel). The Spearman correlation coefficient between mass and density for the current sample of 30 sub-Saturns (excluding K2-60\,b) is equal to $r$ = 0.72. This correlation is comparable to the one for the sample of 23 planets discussed in \citet{2017AJ....153..142P} that is equal to $r$ = 0.79. Notably, almost all of the most massive sub-Saturns from the current sample of 30 planets reside in apparently single-planet systems (blue circles in Figure~\ref{figure-06}). Sub-Saturns in single-planet systems have also often moderate eccentricities, higher than their counterparts in multi-planet systems, as shown in Figure~\ref{figure-07}. As suggested by \cite{2017AJ....153..142P}, the moderate eccentricities of more massive sub-Saturns in apparently single systems and the lack of high-eccentricity, high-mass objects in multi-planet systems may be explained by scattering and merging events during the formation process.

\begin{figure}
\includegraphics[width=\linewidth]{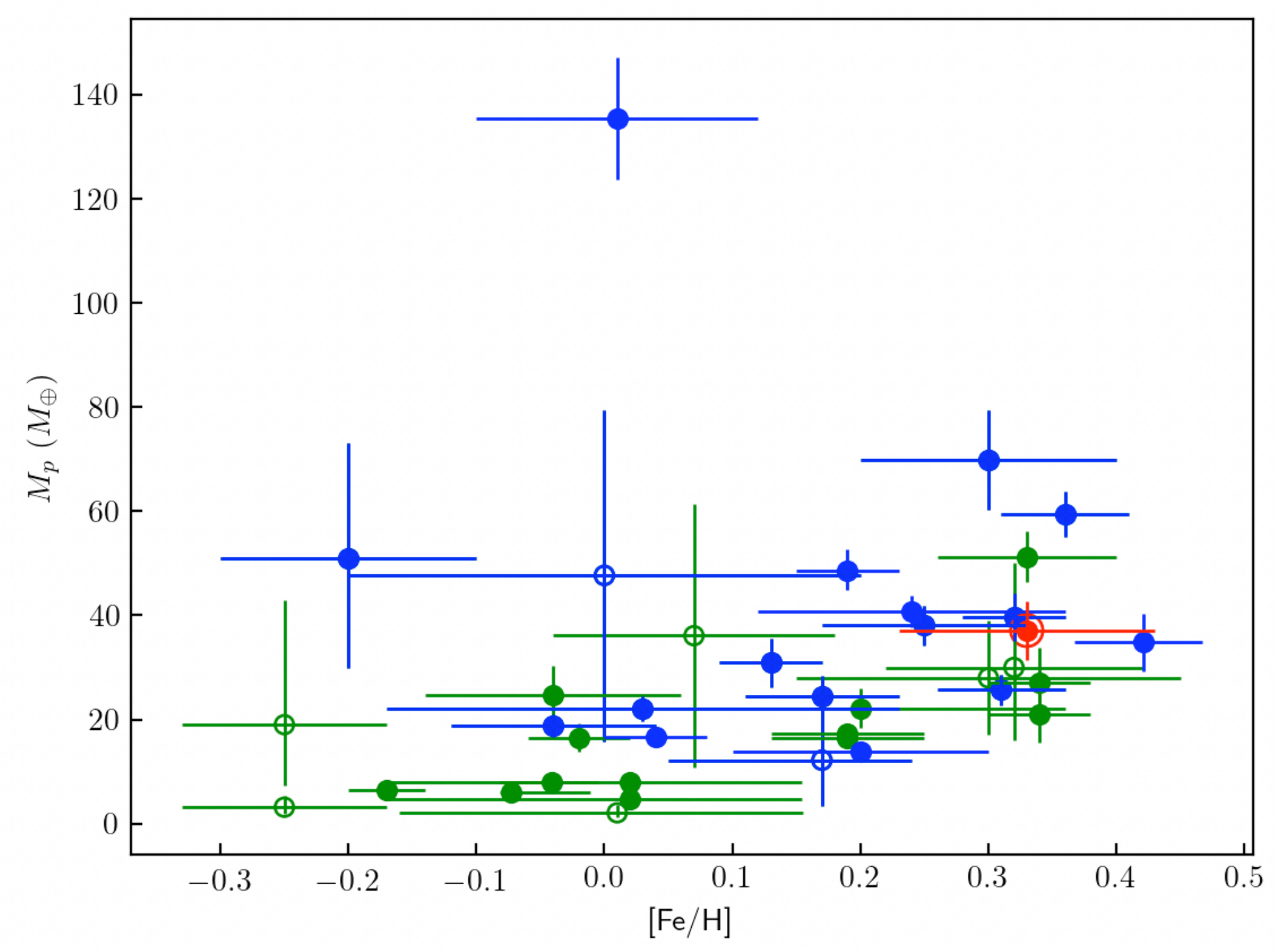}
\caption{Mass of sub-Saturn planets as a function of iron content of their host stars. Samples and point symbols as in Figure~\ref{figure-06}.
\label{figure-08}}
\end{figure}

\cite{2017AJ....153..142P} found a marginal correlation between the stellar metallicity and the mass of sub-Saturn planets (the Spearman correlation coefficient $r$ = 0.57), with the massive sub-Saturns found to orbit metal-rich stars. We confirm this for the current sample of 30 sub-Saturns (excluding K2-60\,b) with exactly the same value of the Spearman correlation coefficient. This is consistent with the results of \cite{2012Natur.486..375B} who, based on the sample of Kepler planets, found that planets larger than $\sim$4\,$\Rea$ orbit stars with relatively high metal content ($-$0.2\,<\,[Fe/H]\,<\,0.5~dex). For the sake of consistency with Figures~\ref{figure-06} and \ref{figure-07}, we included in Figure~\ref{figure-08} the sub-Saturns orbiting binary stars, namely, Kepler-47\,(AB)\,c and d ([Fe/H]\,=\,$-$0.25\,$\pm$\,0.08\,dex) and Kepler-413\,(AB)\,b ([Fe/H]\,=\,$-$0.2\,$\pm$\,0.1\,dex), which were omitted by \cite{2017AJ....153..142P}. Given its mass of $M_\mathrm{p}$\,=\,\mpb\ and the iron content of its host star ($\sfeh$\ =\ \sfehv), \planet\ follows this trend, being a relatively massive sub-Saturn orbiting a metal rich star.

\planet\ has a relatively long orbital period of $\sim$19.9\,days and transits a slightly evolved star in an apparently single-planet system. With an eccentricity of $e$\,=\,\eb, \planet\ is exactly within the range of eccentricities found by \cite{2019AJ....157...61V} for \kepler\ systems with single transiting giant planets ($\ppr$\,>\,6\,$\Rea$). After Kepler-1656\,b \citep{2018AJ....156..147B}, \planet\ is the second most eccentric sub-Saturn known to date. In contrast to Kepler-1656\,b, the mass of $\pbm$\,=\,\mpb, radius of $\pbr$\,=\,\rpb, and mean density of $\pbden$\,=\,\denpb, make \planet\ more similar to HD\,89345\,b \citep[aka K2-234\,b;][]{2018MNRAS.478.4866V,2018AJ....156..127Y}. The moderate eccentricity of \planet\ suggests a formation pathway involving planet-planet gravitational interactions, and make this sub-Saturn planet a member of a relatively rare group of exoplanets and an interesting object for possible future follow-up.

\subsection{Prospects for atmospheric characterisation and Rossiter-McLaughlin effect measurements}
Although \planet\ is a quite puffy planet, the relatively large radius of its host star ($\sr$\,=\,\sradius) results in a quite low transmission signal per scale height ($H$) of the planetary atmosphere (55~ppm). This makes it a difficult target for atmospheric characterisation with current ground- and space-based facilities. The transmission spectroscopy metric (TSM) defined by \cite{2018PASP..130k4401K} for JWST/NIRISS is $\sim$45 for \planet, i.e. two times lower than the threshold TSM for planets with radii $\ppr \in (1.5 - 10.0)$\,$\Rea$ to be selected as high-quality atmospheric characterization targets. The long transit duration ($\sim$8 hours) further complicates ground-based follow-up observations. 

Still, there is a possibility of Rossiter-McLaughlin (RM) effect measurements, for which the overall amplitude is expected to be $\sim$6\,$\mps$, depending on the real values of stellar projected rotation velocity and planetary and stellar radii. With an impact parameter of $b$\,=\,\bb, the transit of \planet\ is close to being central. In such a case the shape of the RM effect would not change significantly with the sky-projected spin-orbit angle $\lambda$, but mainly the RM amplitude, leading to a strong correlation between $\lambda$ and $\svrotsini$ \citep[see, e.g.,][]{2011ApJ...738...50A}. Therefore more precise determination of $\svrotsini$ of this slow rotator, based for instance on the Fourier transform technique \citep[e.g.][and references therein]{1976PASP...88..809S,1990A&A...237..137D,2008oasp.book.....G} applied to single very high resolution and high SNR line profiles, would be needed. Measurements of the sky-projected spin-orbit angle through RM observations may help to test formation scenarios of warm sub-Saturn planets. This gives additional arguments for attempting RM observations, as the probability of a misalignment between the planet's orbital angular momentum vector and its host star's spin axis should be higher if caused by a perturber than by primordial misalignment of the protoplanetary disk. Two full transits of \planet\, observable from the Chilean observatories will occur on July 7$^\mathrm{th}$/8$^\mathrm{th}$ 2020 and August 9$^\mathrm{th}$/10$^\mathrm{th}$ 2021.

\section{Conclusions}
We report here detailed characterisation of a low-density ($\pbden$\,=\,\denpb) sub-Saturn transiting a mildly evolved, metal rich G7 star \host. 
With a mass of $\pbm$\,=\,\mpb, a radius of $\pbr$\,=\,\rpb, and an eccentricity of $e$\,=\,\eb, \planet\ joins the group of sub-Saturns planets in apparently single-planet systems. This second most eccentric sub-Saturn known to date is an interesting object for possible future follow-up observations that may help to test formation scenarios of this intriguing group of planets that are absent in the Solar System.

\section*{Acknowledgements}
This work was supported by the Spanish Ministry of Economy and Competitiveness (MINECO) through grants ESP2016-80435-C2-1-R and ESP2016-80435-C2-2-R. S.M. acknowledges support from the Spanish Ministry with the Ramon y Cajal fellowship number RYC-2015-17697. O.B. acknowledges support from the UK Science and Technology Facilities Council (STFC) under grants ST/S000488/1 and ST/R004846/1. C.M.P., M.F., and I.G. gratefully acknowledge the support of the Swedish National Space Agency (DNR 163/16 and 174/18). J.K., S.G., M.P., S.C., A.P.H., and H.R. acknowledge support by Deutsche Forschungsgemeinschaft (DFG) grants PA525/18-1, PA525/19-1, PA525/20-1, HA 3279/12-1, and RA 714/14-1 within the DFG Schwerpunkt SPP 1992, Exploring the Diversity of Extra-solar Planets. This work is partly supported by JSPS KAKENHI Grant Numbers JP18H01265, JP18H05439 and JP16K17660, and JST PRESTO Grant Number JPMJPR1775. RB acknowledges support from FONDECYT Postdoctoral Fellowship Project 3180246. A.J. acknowledges support from FONDECYT project 1171208. R.B., A.J. and F.R. acknowledge additional support from the Ministry for the Economy, Development, and Tourism's Programa Iniciativa Cient\'{i}fica Milenio through grant IC\,120009, awarded to the Millennium Institute of Astrophysics (MAS).   

We are very grateful to the Gemini-North, NOT, HARPS, and HARPS-N staff members for their unique support during the observations. Based on observations obtained at the Gemini Observatory, which is operated by the Association of Universities for Research in Astronomy, Inc., under a cooperative agreement with the NSF on behalf of the Gemini partnership: the National Science Foundation (United States), National Research Council (Canada), CONICYT (Chile), Ministerio de Ciencia, Tecnolog\'{i}a e Innovaci\'{o}n Productiva (Argentina), Minist\'{e}rio da Ci\^{e}ncia, Tecnologia e Inova\c{c}\~{a}o (Brazil), and Korea Astronomy and Space Science Institute (Republic of Korea). Based on observations obtained with the Nordic Optical Telescope (NOT), operated on the island of La Palma jointly by Denmark, Finland, Iceland, Norway, and Sweden, in the Spanish Observatorio del Roque de los Muchachos (ORM) of the Instituto de Astrof\'isica de Canarias (IAC) under programme 53-109. Based on observations collected at the European Organisation for Astronomical Research in the Southern Hemisphere under ESO programmes 097.C-0571(B), 097.C-0948(A), 098.C-0860(A), 099.C-0491(A), 0101.C-0407(A), and 60.A-9700(G). Based on observations made with the Italian Telescopio Nazionale Galileo (TNG) operated on the island of La Palma by the Fundaci\'on Galileo Galilei of the INAF (Istituto Nazionale di Astrofisica) at the Spanish Observatorio del Roque de los Muchachos of the Instituto de Astrofisica de Canarias under programmes A33TAC\_15, OPT17A\_64, CAT17A\_91, and CAT19A\_97.

This paper includes data collected by the Kepler mission. Funding for the Kepler mission is provided by the NASA Science Mission directorate. This work has made use of data from the European Space Agency (ESA) mission {\it Gaia} (\url{https://www.cosmos.esa.int/gaia}), processed by the {\it Gaia} Data Processing and Analysis Consortium (DPAC, \url{https://www.cosmos.esa.int/web/gaia/dpac/consortium}). Funding for the DPAC has been provided by national institutions, in particular the institutions participating in the {\it Gaia} Multilateral Agreement. This research has made use of NASA Exoplanet Archive and NASA's Astrophysics Data System.

\noindent {\it Facility:} \kepler, \gaia, Gemini-North/NIRI+ALTAIR, ESO/HARPS, TNG/HARPS-N, NOT/FIES.

\noindent {\it Software:} \texttt{BLS}, \texttt{IRAF}, \texttt{PARAM~1.3}, \texttt{pyaneti}, \texttt{SME}, \texttt{SpecMatch-emp}.

\noindent \emph{Data availability:} The data underlying this article are available in the article and in its online supplementary material.


\begin{thebibliography}{}
\makeatletter
\relax
\def\mn@urlcharsother{\let\do\@makeother \do\$\do\&\do\#\do\^\do\_\do\%\do\~}
\def\mn@doi{\begingroup\mn@urlcharsother \@ifnextchar [ {\mn@doi@}
  {\mn@doi@[]}}
\def\mn@doi@[#1]#2{\def\@tempa{#1}\ifx\@tempa\@empty \href
  {http://dx.doi.org/#2} {doi:#2}\else \href {http://dx.doi.org/#2} {#1}\fi
  \endgroup}
\def\mn@eprint#1#2{\mn@eprint@#1:#2::\@nil}
\def\mn@eprint@arXiv#1{\href {http://arxiv.org/abs/#1} {{\tt arXiv:#1}}}
\def\mn@eprint@dblp#1{\href {http://dblp.uni-trier.de/rec/bibtex/#1.xml}
  {dblp:#1}}
\def\mn@eprint@#1:#2:#3:#4\@nil{\def\@tempa {#1}\def\@tempb {#2}\def\@tempc
  {#3}\ifx \@tempc \@empty \let \@tempc \@tempb \let \@tempb \@tempa \fi \ifx
  \@tempb \@empty \def\@tempb {arXiv}\fi \@ifundefined
  {mn@eprint@\@tempb}{\@tempb:\@tempc}{\expandafter \expandafter \csname
  mn@eprint@\@tempb\endcsname \expandafter{\@tempc}}}

\bibitem[\protect\citeauthoryear{{Aigrain}, {Pont}  \& {Zucker}}{{Aigrain}
  et~al.}{2012}]{2012MNRAS.419.3147A}
{Aigrain} S.,  {Pont} F.,   {Zucker} S.,  2012, \mn@doi [\mnras]
  {10.1111/j.1365-2966.2011.19960.x}, \href
  {https://ui.adsabs.harvard.edu/abs/2012MNRAS.419.3147A} {419, 3147}

\bibitem[\protect\citeauthoryear{{Akeson} et~al.,}{{Akeson}
    et~al.}{2013}]{2013PASP..125..989A}
{Akeson} R.~L.,  et~al., 2013, \mn@doi [\pasp] {10.1086/672273}, \href
  {https://ui.adsabs.harvard.edu/abs/2013PASP..125..989A} {125, 989}

\bibitem[\protect\citeauthoryear{{Albrecht} et~al.,}{{Albrecht}
  et~al.}{2011}]{2011ApJ...738...50A}
{Albrecht} S.,  et~al., 2011, \mn@doi [\apj] {10.1088/0004-637X/738/1/50},
  \href {https://ui.adsabs.harvard.edu/abs/2011ApJ...738...50A} {738, 50}

\bibitem[\protect\citeauthoryear{{Awiphan} et~al.,}{{Awiphan}
  et~al.}{2016}]{2016MNRAS.463.2574A}
{Awiphan} S.,  et~al., 2016, \mn@doi [\mnras] {10.1093/mnras/stw2148}, \href
  {https://ui.adsabs.harvard.edu/abs/2016MNRAS.463.2574A} {463, 2574}

\bibitem[\protect\citeauthoryear{{Bailer-Jones}, {Rybizki}, {Fouesneau},
  {Mantelet}  \& {Andrae}}{{Bailer-Jones} et~al.}{2018}]{2018AJ....156...58B}
{Bailer-Jones} C.~A.~L.,  {Rybizki} J.,  {Fouesneau} M.,  {Mantelet} G.,
  {Andrae} R.,  2018, \mn@doi [\aj] {10.3847/1538-3881/aacb21}, \href
  {https://ui.adsabs.harvard.edu/abs/2018AJ....156...58B} {156, 58}

\bibitem[\protect\citeauthoryear{{Baranne} et~al.,}{{Baranne}
  et~al.}{1996}]{Baranne1996}
{Baranne} A.,  et~al., 1996, \aaps, \href
  {https://ui.adsabs.harvard.edu/abs/1996A&AS..119..373B} {119, 373}

\bibitem[\protect\citeauthoryear{{Barrag{\'a}n} \& {Gandolfi}}{{Barrag{\'a}n}
  \& {Gandolfi}}{2017}]{2017ascl.soft06001B}
{Barrag{\'a}n} O.,  {Gandolfi} D.,  2017, {Exotrending: Fast and easy-to-use
  light curve detrending software for exoplanets} (\mn@eprint {ascl}
  {1706.001})

\bibitem[\protect\citeauthoryear{{Barrag{\'a}n} et~al.,}{{Barrag{\'a}n}
  et~al.}{2016}]{2016AJ....152..193B}
{Barrag{\'a}n} O.,  et~al., 2016, \mn@doi [\aj] {10.3847/0004-6256/152/6/193},
  \href {http://adsabs.harvard.edu/abs/2016AJ....152..193B} {152, 193}

\bibitem[\protect\citeauthoryear{{Barrag{\'a}n} et~al.,}{{Barrag{\'a}n}
  et~al.}{2018a}]{2018MNRAS.475.1765B}
{Barrag{\'a}n} O.,  et~al., 2018a, \mn@doi [\mnras] {10.1093/mnras/stx3207},
  \href {https://ui.adsabs.harvard.edu/abs/2018MNRAS.475.1765B} {475, 1765}

\bibitem[\protect\citeauthoryear{{Barrag{\'a}n} et~al.,}{{Barrag{\'a}n}
  et~al.}{2018b}]{2018A&A...612A..95B}
{Barrag{\'a}n} O.,  et~al., 2018b, \mn@doi [\aap]
  {10.1051/0004-6361/201732217}, \href
  {https://ui.adsabs.harvard.edu/abs/2018A&A...612A..95B} {612, A95}

\bibitem[\protect\citeauthoryear{Barrag\'an, Gandolfi  \&
  Antoniciello}{Barrag\'an et~al.}{2019}]{2019MNRAS.482.1017B}
Barrag\'an O.,  Gandolfi D.,   Antoniciello G.,  2019, \mn@doi [\mnras]
  {10.1093/mnras/sty2472}, \href
  {https://ui.adsabs.harvard.edu/#abs/2019MNRAS.482.1017B} {482, 1017}

\bibitem[\protect\citeauthoryear{{Beaug{\'e}} \& {Nesvorn{\'y}}}{{Beaug{\'e}}
  \& {Nesvorn{\'y}}}{2013}]{2013ApJ...763...12B}
{Beaug{\'e}} C.,  {Nesvorn{\'y}} D.,  2013, \mn@doi [\apj]
  {10.1088/0004-637X/763/1/12}, \href
  {https://ui.adsabs.harvard.edu/abs/2013ApJ...763...12B} {763, 12}

\bibitem[\protect\citeauthoryear{{Bonfils} et~al.,}{{Bonfils}
  et~al.}{2012}]{2012A&A...546A..27B}
{Bonfils} X.,  et~al., 2012, \mn@doi [\aap] {10.1051/0004-6361/201219623},
  \href {https://ui.adsabs.harvard.edu/abs/2012A&A...546A..27B} {546, A27}

\bibitem[\protect\citeauthoryear{{Borucki} et~al.,}{{Borucki}
  et~al.}{2010}]{2010Sci...327..977B}
{Borucki} W.~J.,  et~al., 2010, \mn@doi [Science] {10.1126/science.1185402},
  \href {http://adsabs.harvard.edu/abs/2010Sci...327..977B} {327, 977}

\bibitem[\protect\citeauthoryear{{Brady} et~al.,}{{Brady}
  et~al.}{2018}]{2018AJ....156..147B}
{Brady} M.~T.,  et~al., 2018, \mn@doi [\aj] {10.3847/1538-3881/aad773}, \href
  {https://ui.adsabs.harvard.edu/abs/2018AJ....156..147B} {156, 147}

\bibitem[\protect\citeauthoryear{{Bressan}, {Marigo}, {Girardi}, {Salasnich},
  {Dal Cero}, {Rubele}  \& {Nanni}}{{Bressan}
  et~al.}{2012}]{2012MNRAS.427..127B}
{Bressan} A.,  {Marigo} P.,  {Girardi} L.,  {Salasnich} B.,  {Dal Cero} C.,
  {Rubele} S.,   {Nanni} A.,  2012, \mn@doi [\mnras]
  {10.1111/j.1365-2966.2012.21948.x}, \href
  {http://adsabs.harvard.edu/abs/2012MNRAS.427..127B} {427, 127}

\bibitem[\protect\citeauthoryear{{Bruntt} et~al.,}{{Bruntt}
  et~al.}{2010}]{2010MNRAS.405.1907B}
{Bruntt} H.,  et~al., 2010, \mn@doi [\mnras]
  {10.1111/j.1365-2966.2010.16575.x}, \href
  {http://adsabs.harvard.edu/abs/2010MNRAS.405.1907B} {405, 1907}

\bibitem[\protect\citeauthoryear{{Buchhave} et~al.,}{{Buchhave}
  et~al.}{2010}]{2010ApJ...720.1118B}
{Buchhave} L.~A.,  et~al., 2010, \mn@doi [\apj] {10.1088/0004-637X/720/2/1118},
  \href {http://adsabs.harvard.edu/abs/2010ApJ...720.1118B} {720, 1118}

\bibitem[\protect\citeauthoryear{{Buchhave} et~al.,}{{Buchhave}
  et~al.}{2012}]{2012Natur.486..375B}
{Buchhave} L.~A.,  et~al., 2012, \mn@doi [\nat] {10.1038/nature11121}, \href
  {https://ui.adsabs.harvard.edu/abs/2012Natur.486..375B} {486, 375}

\bibitem[\protect\citeauthoryear{Burnham \& Anderson}{Burnham \&
  Anderson}{2002}]{BurnhamAnderson2002}
Burnham K.,  Anderson D.,  2002, Model Selection and Multimodel Inference: A
  Practical Information-Theoretic Approach.
New York: Springer-Verlag

\bibitem[\protect\citeauthoryear{{Christiansen} et~al.,}{{Christiansen}
  et~al.}{2017}]{2017AJ....154..122C}
{Christiansen} J.~L.,  et~al., 2017, \mn@doi [\aj] {10.3847/1538-3881/aa832d},
  \href {https://ui.adsabs.harvard.edu/abs/2017AJ....154..122C} {154, 122}

\bibitem[\protect\citeauthoryear{{Co{\c s}kuno{\v g}lu} et~al.}{{Co{\c
  s}kuno{\v g}lu} et~al.}{2011}]{Coskunoglu11}
{Co{\c s}kuno{\v g}lu} B.,  et~al., 2011, \mn@doi [\mnras]
  {10.1111/j.1365-2966.2010.17983.x}, \href
  {http://adsabs.harvard.edu/abs/2011MNRAS.412.1237C} {412, 1237}

\bibitem[\protect\citeauthoryear{{Cosentino} et~al.,}{{Cosentino}
  et~al.}{2012}]{2012SPIE.8446E..1VC}
{Cosentino} R.,  et~al., 2012, in Ground-based and Airborne Instrumentation for
  Astronomy IV. p. 84461V, \mn@doi{10.1117/12.925738}

\bibitem[\protect\citeauthoryear{{Crossfield} et~al.,}{{Crossfield}
  et~al.}{2016}]{2016ApJS..226....7C}
{Crossfield} I.~J.~M.,  et~al., 2016, \mn@doi [\apjs]
  {10.3847/0067-0049/226/1/7}, \href
  {http://adsabs.harvard.edu/abs/2016ApJS..226....7C} {226, 7}

\bibitem[\protect\citeauthoryear{{Crossfield} et~al.,}{{Crossfield}
  et~al.}{2018}]{2018ApJS..239....5C}
{Crossfield} I.~J.~M.,  et~al., 2018, \mn@doi [\apjs]
  {10.3847/1538-4365/aae155}, \href
  {http://adsabs.harvard.edu/abs/2018ApJS..239....5C} {239, 5}

\bibitem[\protect\citeauthoryear{{Cunha}, {Figueira}, {Santos}, {Lovis}  \&
  {Bou{\'e}}}{{Cunha} et~al.}{2013}]{2013A&A...550A..75C}
{Cunha} D.,  {Figueira} P.,  {Santos} N.~C.,  {Lovis} C.,   {Bou{\'e}} G.,
  2013, \mn@doi [\aap] {10.1051/0004-6361/201220083}, \href
  {https://ui.adsabs.harvard.edu/abs/2013A&A...550A..75C} {550, A75}

\bibitem[\protect\citeauthoryear{{Dai}, {Winn}, {Yu}  \& {Albrecht}}{{Dai}
  et~al.}{2017}]{2017AJ....153...40D}
{Dai} F.,  {Winn} J.~N.,  {Yu} L.,   {Albrecht} S.,  2017, \mn@doi [\aj]
  {10.3847/1538-3881/153/1/40}, \href
  {http://adsabs.harvard.edu/abs/2017AJ....153...40D} {153, 40}

\bibitem[\protect\citeauthoryear{{Demangeon} et~al.,}{{Demangeon}
  et~al.}{2018}]{2018A&A...610A..63D}
{Demangeon} O.~D.~S.,  et~al., 2018, \mn@doi [\aap]
  {10.1051/0004-6361/201731735}, \href
  {https://ui.adsabs.harvard.edu/abs/2018A&A...610A..63D} {610, A63}

\bibitem[\protect\citeauthoryear{{Dorn}, {Khan}, {Heng}, {Connolly}, {Alibert},
  {Benz}  \& {Tackley}}{{Dorn} et~al.}{2015}]{2015A&A...577A..83D}
{Dorn} C.,  {Khan} A.,  {Heng} K.,  {Connolly} J. A.~D.,  {Alibert} Y.,  {Benz}
  W.,   {Tackley} P.,  2015, \mn@doi [\aap] {10.1051/0004-6361/201424915},
  \href {https://ui.adsabs.harvard.edu/abs/2015A&A...577A..83D} {577, A83}

\bibitem[\protect\citeauthoryear{{Dotter}, {Chaboyer}, {Jevremovi{\'c}},
  {Kostov}, {Baron}  \& {Ferguson}}{{Dotter}
  et~al.}{2008}]{2008ApJS..178...89D}
{Dotter} A.,  {Chaboyer} B.,  {Jevremovi{\'c}} D.,  {Kostov} V.,  {Baron} E.,
  {Ferguson} J.~W.,  2008, \mn@doi [\apjs] {10.1086/589654}, \href
  {http://adsabs.harvard.edu/abs/2008ApJS..178...89D} {178, 89}

\bibitem[\protect\citeauthoryear{{Doyle}, {Davies}, {Smalley}, {Chaplin}  \&
  {Elsworth}}{{Doyle} et~al.}{2014}]{2014MNRAS.444.3592D}
{Doyle} A.~P.,  {Davies} G.~R.,  {Smalley} B.,  {Chaplin} W.~J.,   {Elsworth}
  Y.,  2014, \mn@doi [\mnras] {10.1093/mnras/stu1692}, \href
  {http://adsabs.harvard.edu/abs/2014MNRAS.444.3592D} {444, 3592}

\bibitem[\protect\citeauthoryear{{Dravins}, {Lindegren}  \&
  {Torkelsson}}{{Dravins} et~al.}{1990}]{1990A&A...237..137D}
{Dravins} D.,  {Lindegren} L.,   {Torkelsson} U.,  1990, \aap, \href
  {https://ui.adsabs.harvard.edu/abs/1990A&A...237..137D} {237, 137}

\bibitem[\protect\citeauthoryear{{Dressing} et~al.,}{{Dressing}
  et~al.}{2017}]{2017AJ....154..207D}
{Dressing} C.~D.,  et~al., 2017, \mn@doi [\aj] {10.3847/1538-3881/aa89f2},
  \href {http://adsabs.harvard.edu/abs/2017AJ....154..207D} {154, 207}

\bibitem[\protect\citeauthoryear{{Eastman}, {Siverd}  \& {Gaudi}}{{Eastman}
  et~al.}{2010}]{2010PASP..122..935E}
{Eastman} J.,  {Siverd} R.,   {Gaudi} B.~S.,  2010, \mn@doi [\pasp]
  {10.1086/655938}, \href {http://adsabs.harvard.edu/abs/2010PASP..122..935E}
  {122, 935}

\bibitem[\protect\citeauthoryear{{Eigm{\"u}ller} et~al.,}{{Eigm{\"u}ller}
  et~al.}{2017}]{2017AJ....153..130E}
{Eigm{\"u}ller} P.,  et~al., 2017, \mn@doi [\aj] {10.3847/1538-3881/aa5d0b},
  \href {https://ui.adsabs.harvard.edu/abs/2017AJ....153..130E} {153, 130}

\bibitem[\protect\citeauthoryear{{Esposito} et~al.,}{{Esposito}
  et~al.}{2019}]{2019A&A...623A.165E}
{Esposito} M.,  et~al., 2019, \mn@doi [\aap] {10.1051/0004-6361/201834853},
  \href {https://ui.adsabs.harvard.edu/abs/2019A&A...623A.165E} {623, A165}

\bibitem[\protect\citeauthoryear{{Foreman-Mackey}}{{Foreman-Mackey}}{2016}]{2016JOSS....1...24F}
{Foreman-Mackey} D.,  2016, \mn@doi [The Journal of Open Source Software]
  {10.21105/joss.00024}, \href
  {https://ui.adsabs.harvard.edu/abs/2016JOSS....1...24F} {1, 24}

\bibitem[\protect\citeauthoryear{{Fortney}, {Marley}  \& {Barnes}}{{Fortney}
  et~al.}{2007}]{2007ApJ...659.1661F}
{Fortney} J.~J.,  {Marley} M.~S.,   {Barnes} J.~W.,  2007, \mn@doi [\apj]
  {10.1086/512120}, \href
  {https://ui.adsabs.harvard.edu/abs/2007ApJ...659.1661F} {659, 1661}

\bibitem[\protect\citeauthoryear{{Frandsen} \& {Lindberg}}{{Frandsen} \&
  {Lindberg}}{1999}]{1999anot.conf...71F}
{Frandsen} S.,  {Lindberg} B.,  1999, in {Karttunen} H.,  {Piirola} V.,  eds,
  Astrophysics with the NOT. p.~71

\bibitem[\protect\citeauthoryear{{Fridlund} et~al.,}{{Fridlund}
  et~al.}{2017}]{2017A&A...604A..16F}
{Fridlund} M.,  et~al., 2017, \mn@doi [\aap] {10.1051/0004-6361/201730822},
  \href {http://adsabs.harvard.edu/abs/2017A%26A...604A..16F} {604, A16}

\bibitem[\protect\citeauthoryear{{Gaia Collaboration} et~al.,}{{Gaia
  Collaboration} et~al.}{2016}]{2016A&A...595A...1G}
{Gaia Collaboration} et~al., 2016, \mn@doi [\aap]
  {10.1051/0004-6361/201629272}, \href
  {http://adsabs.harvard.edu/abs/2016A%26A...595A...1G} {595, A1}

\bibitem[\protect\citeauthoryear{{Gaia Collaboration} et~al.,}{{Gaia
  Collaboration} et~al.}{2018}]{2018A&A...616A...1G}
{Gaia Collaboration} et~al., 2018, \mn@doi [\aap]
  {10.1051/0004-6361/201833051}, \href
  {http://adsabs.harvard.edu/abs/2018A%26A...616A...1G} {616, A1}

\bibitem[\protect\citeauthoryear{{Gandolfi} et~al.,}{{Gandolfi}
  et~al.}{2008}]{2008ApJ...687.1303G}
{Gandolfi} D.,  et~al., 2008, \mn@doi [\apj] {10.1086/591729}, \href
  {http://adsabs.harvard.edu/abs/2008ApJ...687.1303G} {687, 1303}

\bibitem[\protect\citeauthoryear{{Gandolfi} et~al.,}{{Gandolfi}
  et~al.}{2015}]{2015A&A...576A..11G}
{Gandolfi} D.,  et~al., 2015, \mn@doi [\aap] {10.1051/0004-6361/201425062},
  \href {http://adsabs.harvard.edu/abs/2015A%26A...576A..11G} {576, A11}

\bibitem[\protect\citeauthoryear{{Gandolfi} et~al.,}{{Gandolfi}
  et~al.}{2017}]{2017AJ....154..123G}
{Gandolfi} D.,  et~al., 2017, \mn@doi [\aj] {10.3847/1538-3881/aa832a}, \href
  {https://ui.adsabs.harvard.edu/abs/2017AJ....154..123G} {154, 123}

\bibitem[\protect\citeauthoryear{{Gelman} \& {Rubin}}{{Gelman} \&
  {Rubin}}{1992}]{Gelman1992}
{Gelman} A.,  {Rubin} D.~B.,  1992, \mn@doi [Statistical Science]
  {10.1214/ss/1177011136}, \href
  {http://adsabs.harvard.edu/abs/1992StaSc...7..457G} {7, 457}

\bibitem[\protect\citeauthoryear{{Gray}}{{Gray}}{2008}]{2008oasp.book.....G}
{Gray} D.~F.,  2008, {The Observation and Analysis of Stellar Photospheres}

\bibitem[\protect\citeauthoryear{{Gray} \& {Corbally}}{{Gray} \&
  {Corbally}}{2009}]{2009ssc..book.....G}
{Gray} R.~O.,  {Corbally} J. C.,  2009, {Stellar Spectral Classification}

\bibitem[\protect\citeauthoryear{{Hekker}, {Reffert}, {Quirrenbach},
  {Mitchell}, {Fischer}, {Marcy}  \& {Butler}}{{Hekker}
  et~al.}{2006}]{2006A&A...454..943H}
{Hekker} S.,  {Reffert} S.,  {Quirrenbach} A.,  {Mitchell} D.~S.,  {Fischer}
  D.~A.,  {Marcy} G.~W.,   {Butler} R.~P.,  2006, \mn@doi [\aap]
  {10.1051/0004-6361:20064946}, \href
  {https://ui.adsabs.harvard.edu/abs/2006A&A...454..943H} {454, 943}

\bibitem[\protect\citeauthoryear{{Hekker}, {Snellen}, {Aerts}, {Quirrenbach},
  {Reffert}  \& {Mitchell}}{{Hekker} et~al.}{2008}]{2008A&A...480..215H}
{Hekker} S.,  {Snellen} I.~A.~G.,  {Aerts} C.,  {Quirrenbach} A.,  {Reffert}
  S.,   {Mitchell} D.~S.,  2008, \mn@doi [\aap] {10.1051/0004-6361:20078321},
  \href {https://ui.adsabs.harvard.edu/abs/2008A&A...480..215H} {480, 215}

\bibitem[\protect\citeauthoryear{{Hirano} et~al.,}{{Hirano}
  et~al.}{2018}]{2018AJ....155..127H}
{Hirano} T.,  et~al., 2018, \mn@doi [\aj] {10.3847/1538-3881/aaa9c1}, \href
  {https://ui.adsabs.harvard.edu/abs/2018AJ....155..127H} {155, 127}

\bibitem[\protect\citeauthoryear{{Hjorth} et~al.,}{{Hjorth}
  et~al.}{2019}]{2019MNRAS.484.3522H}
{Hjorth} M.,  et~al., 2019, \mn@doi [\mnras] {10.1093/mnras/stz139}, \href
  {https://ui.adsabs.harvard.edu/abs/2019MNRAS.484.3522H} {484, 3522}

\bibitem[\protect\citeauthoryear{{Hodapp} et~al.,}{{Hodapp}
  et~al.}{2003}]{2003PASP..115.1388H}
{Hodapp} K.~W.,  et~al., 2003, \mn@doi [\pasp] {10.1086/379669}, \href
  {http://adsabs.harvard.edu/abs/2003PASP..115.1388H} {115, 1388}

\bibitem[\protect\citeauthoryear{{Howell} et~al.,}{{Howell}
  et~al.}{2014}]{2014PASP..126..398H}
{Howell} S.~B.,  et~al., 2014, \mn@doi [\pasp] {10.1086/676406}, \href
  {http://adsabs.harvard.edu/abs/2014PASP..126..398H} {126, 398}

\bibitem[\protect\citeauthoryear{{Jenkins} et~al.,}{{Jenkins}
  et~al.}{2010}]{2010ApJ...713L..87J}
{Jenkins} J.~M.,  et~al., 2010, \mn@doi [\apjl] {10.1088/2041-8205/713/2/L87},
  \href {http://adsabs.harvard.edu/abs/2010ApJ...713L..87J} {713, L87}

\bibitem[\protect\citeauthoryear{{Johnson} \& {Soderblom}}{{Johnson} \&
  {Soderblom}}{1987}]{JohnsonSoderblom87}
{Johnson} D.~R.~H.,  {Soderblom} D.~R.,  1987, \mn@doi [\aj] {10.1086/114370},
  \href {http://adsabs.harvard.edu/abs/1987AJ.....93..864J} {93, 864}

\bibitem[\protect\citeauthoryear{{Kempton} et~al.,}{{Kempton}
  et~al.}{2018}]{2018PASP..130k4401K}
{Kempton} E. M.~R.,  et~al., 2018, \mn@doi [\pasp] {10.1088/1538-3873/aadf6f},
  \href {https://ui.adsabs.harvard.edu/abs/2018PASP..130k4401K} {130, 114401}

\bibitem[\protect\citeauthoryear{{Kipping}}{{Kipping}}{2010}]{2010MNRAS.408.1758K}
{Kipping} D.~M.,  2010, \mn@doi [\mnras] {10.1111/j.1365-2966.2010.17242.x},
  \href {http://adsabs.harvard.edu/abs/2010MNRAS.408.1758K} {408, 1758}

\bibitem[\protect\citeauthoryear{{Kov{\'a}cs}, {Zucker}  \&
  {Mazeh}}{{Kov{\'a}cs} et~al.}{2002}]{2002A&A...391..369K}
{Kov{\'a}cs} G.,  {Zucker} S.,   {Mazeh} T.,  2002, \mn@doi [\aap]
  {10.1051/0004-6361:20020802}, \href
  {http://adsabs.harvard.edu/abs/2002A%26A...391..369K} {391, 369}

\bibitem[\protect\citeauthoryear{{Kurucz}}{{Kurucz}}{2013}]{2013ascl.soft03024K}
{Kurucz} R.~L.,  2013, {ATLAS12: Opacity sampling model atmosphere program},
  Astrophysics Source Code Library (\mn@eprint {ascl} {1303.024})

\bibitem[\protect\citeauthoryear{{Laughlin}, {Crismani}  \& {Adams}}{{Laughlin}
  et~al.}{2011}]{2011ApJ...729L...7L}
{Laughlin} G.,  {Crismani} M.,   {Adams} F.~C.,  2011, \mn@doi [\apjl]
  {10.1088/2041-8205/729/1/L7}, \href
  {https://ui.adsabs.harvard.edu/abs/2011ApJ...729L...7L} {729, L7}

\bibitem[\protect\citeauthoryear{{Livingston} et~al.,}{{Livingston}
  et~al.}{2018}]{2018AJ....156..277L}
{Livingston} J.~H.,  et~al., 2018, \mn@doi [\aj] {10.3847/1538-3881/aae778},
  \href {http://adsabs.harvard.edu/abs/2018AJ....156..277L} {156, 277}

\bibitem[\protect\citeauthoryear{{Luri} et~al.,}{{Luri}
  et~al.}{2018}]{2018A&A...616A...9L}
{Luri} X.,  et~al., 2018, \mn@doi [\aap] {10.1051/0004-6361/201832964}, \href
  {http://adsabs.harvard.edu/abs/2018A%26A...616A...9L} {616, A9}

\bibitem[\protect\citeauthoryear{{Malavolta} et~al.,}{{Malavolta}
  et~al.}{2018}]{2018AJ....155..107M}
{Malavolta} L.,  et~al., 2018, \mn@doi [\aj] {10.3847/1538-3881/aaa5b5}, \href
  {https://ui.adsabs.harvard.edu/abs/2018AJ....155..107M} {155, 107}

\bibitem[\protect\citeauthoryear{{Mandel} \& {Agol}}{{Mandel} \&
  {Agol}}{2002}]{2002ApJ...580L.171M}
{Mandel} K.,  {Agol} E.,  2002, \mn@doi [\apjl] {10.1086/345520}, \href
  {http://adsabs.harvard.edu/abs/2002ApJ...580L.171M} {580, L171}

\bibitem[\protect\citeauthoryear{{Mayer}, {Wadsley}, {Quinn}  \&
  {Stadel}}{{Mayer} et~al.}{2005}]{2005MNRAS.363..641M}
{Mayer} L.,  {Wadsley} J.,  {Quinn} T.,   {Stadel} J.,  2005, \mn@doi [\mnras]
  {10.1111/j.1365-2966.2005.09468.x}, \href
  {https://ui.adsabs.harvard.edu/abs/2005MNRAS.363..641M} {363, 641}

\bibitem[\protect\citeauthoryear{{Mayo} et~al.,}{{Mayo}
  et~al.}{2018}]{2018AJ....155..136M}
{Mayo} A.~W.,  et~al., 2018, \mn@doi [\aj] {10.3847/1538-3881/aaadff}, \href
  {http://adsabs.harvard.edu/abs/2018AJ....155..136M} {155, 136}

\bibitem[\protect\citeauthoryear{{Mayor} et~al.,}{{Mayor}
  et~al.}{2003}]{2003Msngr.114...20M}
{Mayor} M.,  et~al., 2003, The Messenger, \href
  {http://adsabs.harvard.edu/abs/2003Msngr.114...20M} {114, 20}

\bibitem[\protect\citeauthoryear{{Mazeh}, {Holczer}  \& {Faigler}}{{Mazeh}
  et~al.}{2016}]{2016A&A...589A..75M}
{Mazeh} T.,  {Holczer} T.,   {Faigler} S.,  2016, \mn@doi [\aap]
  {10.1051/0004-6361/201528065}, \href
  {https://ui.adsabs.harvard.edu/abs/2016A&A...589A..75M} {589, A75}

\bibitem[\protect\citeauthoryear{{Montet} et~al.,}{{Montet}
  et~al.}{2015}]{2015ApJ...809...25M}
{Montet} B.~T.,  et~al., 2015, \mn@doi [\apj] {10.1088/0004-637X/809/1/25},
  \href {http://adsabs.harvard.edu/abs/2015ApJ...809...25M} {809, 25}

\bibitem[\protect\citeauthoryear{{Nelson}}{{Nelson}}{2000}]{2000ApJ...537L..65N}
{Nelson} A.~F.,  2000, \mn@doi [\apj] {10.1086/312752}, \href
  {https://ui.adsabs.harvard.edu/abs/2000ApJ...537L..65N} {537, L65}

\bibitem[\protect\citeauthoryear{{Ofir}}{{Ofir}}{2014}]{2014A&A...561A.138O}
{Ofir} A.,  2014, \mn@doi [\aap] {10.1051/0004-6361/201220860}, \href
  {http://adsabs.harvard.edu/abs/2014A%26A...561A.138O} {561, A138}

\bibitem[\protect\citeauthoryear{{Pepe} et~al.,}{{Pepe}
  et~al.}{2013}]{2013Natur.503..377P}
{Pepe} F.,  et~al., 2013, \mn@doi [\nat] {10.1038/nature12768}, \href
  {http://adsabs.harvard.edu/abs/2013Natur.503..377P} {503, 377}

\bibitem[\protect\citeauthoryear{{Persson} et~al.,}{{Persson}
  et~al.}{2018}]{2018A&A...618A..33P}
{Persson} C.~M.,  et~al., 2018, \mn@doi [\aap] {10.1051/0004-6361/201832867},
  \href {http://adsabs.harvard.edu/abs/2018A%26A...618A..33P} {618, A33}

\bibitem[\protect\citeauthoryear{{Petigura} et~al.,}{{Petigura}
  et~al.}{2016}]{2016ApJ...818...36P}
{Petigura} E.~A.,  et~al., 2016, \mn@doi [\apj] {10.3847/0004-637X/818/1/36},
  \href {https://ui.adsabs.harvard.edu/abs/2016ApJ...818...36P} {818, 36}

\bibitem[\protect\citeauthoryear{{Petigura} et~al.,}{{Petigura}
  et~al.}{2017}]{2017AJ....153..142P}
{Petigura} E.~A.,  et~al., 2017, \mn@doi [\aj] {10.3847/1538-3881/aa5ea5},
  \href {https://ui.adsabs.harvard.edu/abs/2017AJ....153..142P} {153, 142}

\bibitem[\protect\citeauthoryear{{Petigura} et~al.,}{{Petigura}
  et~al.}{2018}]{2018AJ....155...21P}
{Petigura} E.~A.,  et~al., 2018, \mn@doi [\aj] {10.3847/1538-3881/aa9b83},
  \href {https://ui.adsabs.harvard.edu/abs/2018AJ....155...21P} {155, 21}

\bibitem[\protect\citeauthoryear{{Piskunov} \& {Valenti}}{{Piskunov} \&
  {Valenti}}{2017}]{2017A&A...597A..16P}
{Piskunov} N.,  {Valenti} J.~A.,  2017, \mn@doi [\aap]
  {10.1051/0004-6361/201629124}, \href
  {https://ui.adsabs.harvard.edu/abs/2017A&A...597A..16P} {597, A16}

\bibitem[\protect\citeauthoryear{{Pollacco} et~al.,}{{Pollacco}
  et~al.}{2006}]{2006PASP..118.1407P}
{Pollacco} D.~L.,  et~al., 2006, \mn@doi [\pasp] {10.1086/508556}, \href
  {https://ui.adsabs.harvard.edu/abs/2006PASP..118.1407P} {118, 1407}

\bibitem[\protect\citeauthoryear{{Prieto-Arranz} et~al.,}{{Prieto-Arranz}
  et~al.}{2018}]{2018A&A...618A.116P}
{Prieto-Arranz} J.,  et~al., 2018, \mn@doi [\aap]
  {10.1051/0004-6361/201832872}, \href
  {https://ui.adsabs.harvard.edu/abs/2018A&A...618A.116P} {618, A116}

\bibitem[\protect\citeauthoryear{{Reddy}, {Lambert}  \& {Allende
  Prieto}}{{Reddy} et~al.}{2006}]{Reddy06}
{Reddy} B.~E.,  {Lambert} D.~L.,   {Allende Prieto} C.,  2006, \mn@doi [\mnras]
  {10.1111/j.1365-2966.2006.10148.x}, \href
  {http://adsabs.harvard.edu/abs/2006MNRAS.367.1329R} {367, 1329}

\bibitem[\protect\citeauthoryear{{Robin}, {Reyl{\'e}}, {Derri{\`e}re}  \&
  {Picaud}}{{Robin} et~al.}{2003}]{2003A&A...409..523R}
{Robin} A.~C.,  {Reyl{\'e}} C.,  {Derri{\`e}re} S.,   {Picaud} S.,  2003,
  \mn@doi [\aap] {10.1051/0004-6361:20031117}, \href
  {https://ui.adsabs.harvard.edu/abs/2003A&A...409..523R} {409, 523}

\bibitem[\protect\citeauthoryear{{Rodriguez}, {Vanderburg}, {Eastman}, {Mann},
  {Crossfield}, {Ciardi}, {Latham}  \& {Quinn}}{{Rodriguez}
  et~al.}{2018}]{2018AJ....155...72R}
{Rodriguez} J.~E.,  {Vanderburg} A.,  {Eastman} J.~D.,  {Mann} A.~W.,
  {Crossfield} I. J.~M.,  {Ciardi} D.~R.,  {Latham} D.~W.,   {Quinn} S.~N.,
  2018, \mn@doi [\aj] {10.3847/1538-3881/aaa292}, \href
  {https://ui.adsabs.harvard.edu/abs/2018AJ....155...72R} {155, 72}

\bibitem[\protect\citeauthoryear{{Schmitt} et~al.,}{{Schmitt}
  et~al.}{2016}]{2016AJ....151..159S}
{Schmitt} J.~R.,  et~al., 2016, \mn@doi [\aj] {10.3847/0004-6256/151/6/159},
  \href {https://ui.adsabs.harvard.edu/abs/2016AJ....151..159S} {151, 159}

\bibitem[\protect\citeauthoryear{{Smith} \& {Gray}}{{Smith} \&
  {Gray}}{1976}]{1976PASP...88..809S}
{Smith} M.~A.,  {Gray} D.~F.,  1976, \mn@doi [\pasp] {10.1086/130029}, \href
  {https://ui.adsabs.harvard.edu/abs/1976PASP...88..809S} {88, 809}

\bibitem[\protect\citeauthoryear{{Smith} \& {WASP Consortium}}{{Smith} \& {WASP
  Consortium}}{2014}]{2014CoSka..43..500S}
{Smith} A.~M.~S.,  {WASP Consortium} 2014, Contributions of the Astronomical
  Observatory Skalnate Pleso, \href
  {https://ui.adsabs.harvard.edu/abs/2014CoSka..43..500S} {43, 500}

\bibitem[\protect\citeauthoryear{{Southworth}, {Wheatley}  \&
  {Sams}}{{Southworth} et~al.}{2007}]{Sotuhworth2007}
{Southworth} J.,  {Wheatley} P.~J.,   {Sams} G.,  2007, \mn@doi [\mnras]
  {10.1111/j.1745-3933.2007.00324.x}, \href
  {https://ui.adsabs.harvard.edu/abs/2007MNRAS.379L..11S} {379, L11}

\bibitem[\protect\citeauthoryear{{Szab{\'o}} \& {Kiss}}{{Szab{\'o}} \&
  {Kiss}}{2011}]{2011ApJ...727L..44S}
{Szab{\'o}} G.~M.,  {Kiss} L.~L.,  2011, \mn@doi [\apjl]
  {10.1088/2041-8205/727/2/L44}, \href
  {https://ui.adsabs.harvard.edu/abs/2011ApJ...727L..44S} {727, L44}

\bibitem[\protect\citeauthoryear{{Tayar}, {Stassun}  \& {Corsaro}}{{Tayar}
  et~al.}{2019}]{2019ApJ...883..195T}
{Tayar} J.,  {Stassun} K.~G.,   {Corsaro} E.,  2019, \mn@doi [\apj]
  {10.3847/1538-4357/ab3db1}, \href
  {https://ui.adsabs.harvard.edu/abs/2019ApJ...883..195T} {883, 195}

\bibitem[\protect\citeauthoryear{{Telting} et~al.,}{{Telting}
  et~al.}{2014}]{2014AN....335...41T}
{Telting} J.~H.,  et~al., 2014, \mn@doi [Astronomische Nachrichten]
  {10.1002/asna.201312007}, \href
  {http://adsabs.harvard.edu/abs/2014AN....335...41T} {335, 41}

\bibitem[\protect\citeauthoryear{{Th{\'e}bault}, {Marzari}  \&
  {Scholl}}{{Th{\'e}bault} et~al.}{2006}]{2006Icar..183..193T}
{Th{\'e}bault} P.,  {Marzari} F.,   {Scholl} H.,  2006, \mn@doi [\icarus]
  {10.1016/j.icarus.2006.01.022}, \href
  {https://ui.adsabs.harvard.edu/abs/2006Icar..183..193T} {183, 193}

\bibitem[\protect\citeauthoryear{{Torres}, {Andersen}  \&
  {Gim{\'e}nez}}{{Torres} et~al.}{2010}]{2010A&ARv..18...67T}
{Torres} G.,  {Andersen} J.,   {Gim{\'e}nez} A.,  2010, \mn@doi [\aapr]
  {10.1007/s00159-009-0025-1}, \href
  {http://adsabs.harvard.edu/abs/2010A%26ARv..18...67T} {18, 67}

\bibitem[\protect\citeauthoryear{{Udry}, {Mayor}  \& {Queloz}}{{Udry}
  et~al.}{1999}]{1999ASPC..185..367U}
{Udry} S.,  {Mayor} M.,   {Queloz} D.,  1999, in {Hearnshaw} J.~B.,  {Scarfe}
  C.~D.,  eds,  Astronomical Society of the Pacific Conference Series Vol. 185,
  IAU Colloq. 170: Precise Stellar Radial Velocities. p.~367

\bibitem[\protect\citeauthoryear{{Valenti} \& {Fischer}}{{Valenti} \&
  {Fischer}}{2005}]{2005ApJS..159..141V}
{Valenti} J.~A.,  {Fischer} D.~A.,  2005, \mn@doi [\apjs] {10.1086/430500},
  \href {http://adsabs.harvard.edu/abs/2005ApJS..159..141V} {159, 141}

\bibitem[\protect\citeauthoryear{{Valenti} \& {Piskunov}}{{Valenti} \&
  {Piskunov}}{1996}]{1996A&AS..118..595V}
{Valenti} J.~A.,  {Piskunov} N.,  1996, \aaps, \href
  {http://adsabs.harvard.edu/abs/1996A%26AS..118..595V} {118, 595}

\bibitem[\protect\citeauthoryear{{Van Eylen} \& {Albrecht}}{{Van Eylen} \&
  {Albrecht}}{2015}]{2015ApJ...808..126V}
{Van Eylen} V.,  {Albrecht} S.,  2015, \mn@doi [\apj]
  {10.1088/0004-637X/808/2/126}, \href
  {http://adsabs.harvard.edu/abs/2015ApJ...808..126V} {808, 126}

\bibitem[\protect\citeauthoryear{{Van Eylen} et~al.,}{{Van Eylen}
  et~al.}{2018}]{2018MNRAS.478.4866V}
{Van Eylen} V.,  et~al., 2018, \mn@doi [\mnras] {10.1093/mnras/sty1390}, \href
  {https://ui.adsabs.harvard.edu/abs/2018MNRAS.478.4866V} {478, 4866}

\bibitem[\protect\citeauthoryear{{Van Eylen} et~al.,}{{Van Eylen}
  et~al.}{2019}]{2019AJ....157...61V}
{Van Eylen} V.,  et~al., 2019, \mn@doi [\aj] {10.3847/1538-3881/aaf22f}, \href
  {https://ui.adsabs.harvard.edu/abs/2019AJ....157...61V} {157, 61}

\bibitem[\protect\citeauthoryear{{Vanderburg} et~al.,}{{Vanderburg}
  et~al.}{2015}]{2015ApJ...800...59V}
{Vanderburg} A.,  et~al., 2015, \mn@doi [\apj] {10.1088/0004-637X/800/1/59},
  \href {https://ui.adsabs.harvard.edu/abs/2015ApJ...800...59V} {800, 59}

\bibitem[\protect\citeauthoryear{{Vanderburg} et~al.,}{{Vanderburg}
  et~al.}{2016}]{2016ApJS..222...14V}
{Vanderburg} A.,  et~al., 2016, \mn@doi [\apjs] {10.3847/0067-0049/222/1/14},
  \href {http://adsabs.harvard.edu/abs/2016ApJS..222...14V} {222, 14}

\bibitem[\protect\citeauthoryear{{Vogt} et~al.,}{{Vogt}
  et~al.}{1994}]{1994SPIE.2198..362V}
{Vogt} S.~S.,  et~al., 1994, in {Crawford} D.~L.,  {Craine} E.~R.,  eds,
  Society of Photo-Optical Instrumentation Engineers (SPIE) Conference Series
  Vol. 2198, \procspie. p.~362, \mn@doi{10.1117/12.176725}

\bibitem[\protect\citeauthoryear{{Winn}, {Sanchis-Ojeda}  \&
  {Rappaport}}{{Winn} et~al.}{2018}]{2018NewAR..83...37W}
{Winn} J.~N.,  {Sanchis-Ojeda} R.,   {Rappaport} S.,  2018, \mn@doi [\nar]
  {10.1016/j.newar.2019.03.006}, \href
  {https://ui.adsabs.harvard.edu/abs/2018NewAR..83...37W} {83, 37}

\bibitem[\protect\citeauthoryear{{Yee}, {Petigura}  \& {von Braun}}{{Yee}
  et~al.}{2017}]{2017ApJ...836...77Y}
{Yee} S.~W.,  {Petigura} E.~A.,   {von Braun} K.,  2017, \mn@doi [\apj]
  {10.3847/1538-4357/836/1/77}, \href
  {http://adsabs.harvard.edu/abs/2017ApJ...836...77Y} {836, 77}

\bibitem[\protect\citeauthoryear{{Yu} et~al.,}{{Yu}
  et~al.}{2018a}]{2018AJ....156...22Y}
{Yu} L.,  et~al., 2018a, \mn@doi [\aj] {10.3847/1538-3881/aac6e6}, \href
  {http://adsabs.harvard.edu/abs/2018AJ....156...22Y} {156, 22}

\bibitem[\protect\citeauthoryear{{Yu} et~al.,}{{Yu}
  et~al.}{2018b}]{2018AJ....156..127Y}
{Yu} L.,  et~al., 2018b, \mn@doi [\aj] {10.3847/1538-3881/aad6e7}, \href
  {https://ui.adsabs.harvard.edu/abs/2018AJ....156..127Y} {156, 127}

\bibitem[\protect\citeauthoryear{{Zechmeister} \& {K{\"u}rster}}{{Zechmeister}
  \& {K{\"u}rster}}{2009}]{2009A&A...496..577Z}
{Zechmeister} M.,  {K{\"u}rster} M.,  2009, \mn@doi [\aap]
  {10.1051/0004-6361:200811296}, \href
  {https://ui.adsabs.harvard.edu/abs/2009A&A...496..577Z} {496, 577}

\makeatother
\end{thebibliography}

\noindent$^{1}$Instituto de Astrof\'\i sica de Canarias (IAC), 38205 La Laguna, Tenerife, Spain\\
$^{2}$Departamento de Astrof\'\i sica, Universidad de La Laguna (ULL), 38206 La Laguna, Tenerife, Spain\\
$^{3}$Dipartimento di Fisica, Universit\'a di Torino, Via P. Giuria 1, I-10125, Torino, Italy\\
$^{4}$Department of Earth and Planetary Sciences, Tokyo Institute of Technology, 2-12-1 Ookayama, Meguro-ku, Tokio, Japan\\
$^{5}$Sub-department of Astrophysics, Department of Physics, University of Oxford, Oxford OX1 3RH, UK\\
$^{6}$Astrobiology Center, NINS, 2-21-1 Osawa, Mitaka, Tokyo 181-8588, Japan\\
$^{7}$National Astronomical Observatory of Japan, NINS, 2-21-1 Osawa, Mitaka, Tokyo 181-8588, Japan\\
$^{8}$Department of Astrophysical Sciences, Princeton University, 4 Ivy Lane, Princeton, NJ, 08544, USA\\
$^{9}$Department of Physics and Kavli Institute for Astrophysics and Space Research, MIT, Cambridge, MA 02139, USA\\
$^{10}${Chalmers University of Technology, Department of Space, Earth and Environment, Onsala Space Observatory,  SE-439 92 Onsala, Sweden}\\
$^{11}$Leiden Observatory, University of Leiden, PO Box 9513, 2300 RA, Leiden, The Netherlands\\
$^{12}$Las Cumbres Observatory, 6740 Cortona Dr., Ste. 102, Goleta, CA 93117 USA\\
$^{13}$Rheinisches Institut f\"ur Umweltforschung an der Universit\"at zu K\"oln, Aachener Strasse 209, 50931 K\"oln, Germany\\
$^{14}$Department of Astronomy, The University of Tokyo, 7-3-1 Hongo, Bunkyo-ku, Tokyo 113-0033, Japan\\
$^{15}$Th\"uringer Landessternwarte Tautenburg,  D-07778 Tautenburg, Germany\\
$^{16}$Stellar Astrophysics Centre, Department of Physics and Astronomy, Aarhus University, Ny Munkegade 120, DK-8000 Aarhus C\\
$^{17}$Center for Astronomy and Astrophysics, TU Berlin, Hardenbergstr. 36, 10623 Berlin, Germany\\
$^{18}$Department of Astronomy and McDonald Observatory, University of Texas at Austin, 2515 Speedway, Austin, TX 78712, USA\\
$^{19}$Institute of Planetary Research, German Aerospace Center (DLR), Rutherfordstrasse 2, D-12489 Berlin, Germany\\
$^{20}$Astronomical Institute, Czech Academy of Sciences, Fri\v{c}ova 298, 25165, Ond\v{r}ejov, Czech Republic\\
${21}$Valencian International University (VIU), Pintor Sorolla 21, 46002 Valencia, Spain\\
$^{22}$ Center of Astro-Engineering UC, Pontificia Universidad Cat\'olica de Chile,\\
$^{23}$Millennium Institute for Astrophysics, Av.\ Vicu\~{n}a Mackenna 4860, 782-0436 Macul, Santiago, Chile\\
$^{24}$Instituto de Astrof\'{i}sica, Pontificia Universidad Cat\'{o}lica de Chile, Av. Vicu\~{n}a Mackenna 4860, 782-0436 Macul, Santiago, Chile\\
$^{25}$Facultad de Ingenier\'ia y Ciencias, Universidad Adolfo Ib\'a\~nez, Av.\ Diagonal las Torres 2640, Pe\~nalol\'en, Santiago, Chile\\
${26}$Max-Planck-Institut f\"ur Astronomie, K\"onigstuhl 17, 69117 Heidelberg, Germany\\
$^{27}$European Southern Observatory, Alonso de C\'ordova 3107, Vitacura, Casilla 19001, Santiago de Chile, Chile\\
$^{28}$Astronomy Department and Van Vleck Observatory, Wesleyan University, Middletown, CT 06459, USA\\
$^{29}$JST, PRESTO, 2-21-1 Osawa, Mitaka, Tokyo 181-8588, Japan\\
$^{30}$Department of Earth, Atmospheric and Planetary Sciences, MIT, 77 Massachusetts Avenue, Cambridge, MA 02139, USA\\
$^{31}$Institut de Ci\`encies de l'Espai (ICE, CSIC), Campus UAB,C/ de Can Magrans s/n, E-08193 Bellaterra, Spain\\
$^{32}$Institut d'Estudis Espacials de Catalunya (IEEC), C/ Gran Capit\'a 2-4, E-08034 Barcelona, Spain\\
$^{33}$Department of Theoretical Physics and Astrophysics, Masaryk University, Kotl\'{a}\v{r}sk\'{a} 2, 61137 Brno, Czech Republic\\
$^{34}$Astronomical Institute, Faculty of Mathematics and Physics, Charles University, V Hole\v{s}ovi{\v c}k\'ach 2, 180 00, Praha, Czech Republic\\

\appendix
\section{Corner plot for fitted parameters}
\begin{figure*}
\centering
\includegraphics[width=\textwidth]{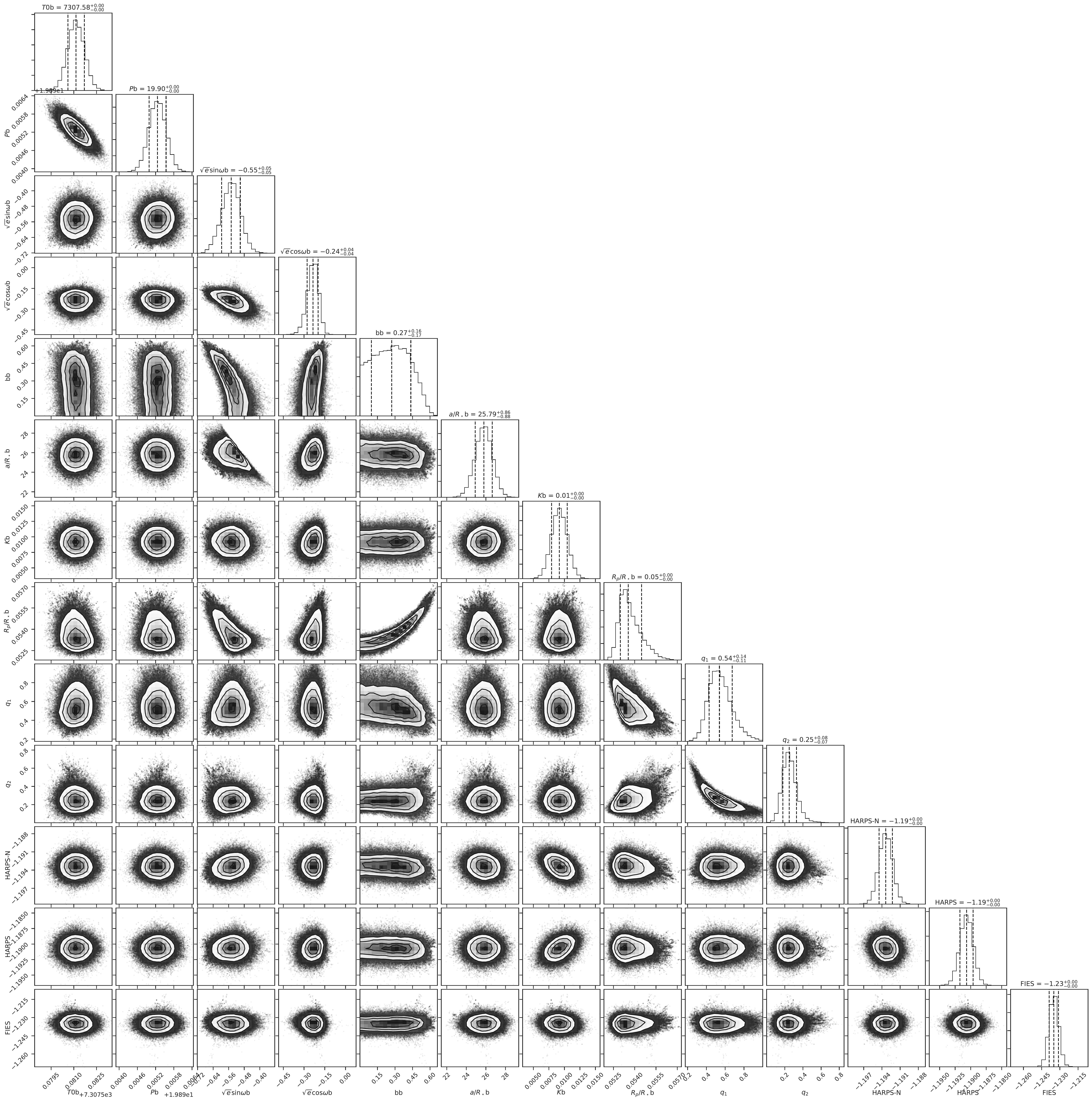}
\caption{Corner plot for the fitted parameters of the K2-280 system. This figure was created using \texttt{corner.py} \citep{2016JOSS....1...24F}.}
\label{figure-09}
\end{figure*}{}

\bsp
\label{lastpage}
\end{document}